\documentclass[a4paper,fleqn]{cas-sc}
\usepackage[authoryear]{natbib}
\usepackage[T1]{fontenc}
\usepackage{graphicx}
\usepackage{pifont}
\usepackage{latexsym}
\usepackage{datetime}
\usepackage{hyperref}
\usepackage{adjustbox}
\usepackage{color,soul}
\usepackage{enumitem}

\newdate{date}{01}{05}{2022}
\newdate{date1}{15}{02}{2022}

\usepackage{url}
\usepackage{caption}
\usepackage{underscore}
\usepackage{graphicx}
\usepackage{subcaption}
\usepackage{gensymb}
\usepackage[english]{babel}
\usepackage{subfiles}
\usepackage{listings}
\usepackage{tabularx}
\usepackage{todonotes}
\usepackage{pdflscape}
\usepackage{multirow}
\usepackage{footnote}

\definecolor{pastelred}{rgb}{1.0, 0.41, 0.38}

\newcommand{\hlt}[1]{\textcolor{blue}{#1}}
\renewcommand{\hlt}[1]{\textnormal{#1}}
\renewcommand{\vec}[1]{\mathbf{#1}}

\newcommand{\PWC}{PwC}
\newcommand{\CG}{SciBERT}
\newcommand{\Factored}{SciBERT-DFG}
\newcommand{\CGBinary}{SciBERT-DFGB}
\newcommand{\FG}{SciBERT-LFG}
\newcommand{\FGBinary}{SciBERT-LFGB}
\newcommand{\FLRCG}{SciBERT\textsubscript{CRF}}
\newcommand{\FLRFactored}{SciBERT\textsubscript{CRF}-DFG}
\newcommand{\FLRCGBinary}{SciBERT\textsubscript{CRF}-DFGB}
\newcommand{\FLRFG}{SciBERT\textsubscript{CRF}-LFG}
\newcommand{\FLRFGBinary}{SciBERT\textsubscript{CRF}-LFGB}
\newcommand{\blstmglove}{BiLSTM\textsubscript{CRF}}
\newcommand{\blstm}{BiLSTM}
\newcommand{\scrx}{SciREX}

\newcommand{\uls}{\begin{itemize}[leftmargin=*]}
\newcommand{\ule}{\end{itemize}}
\newcommand{\ols}{\begin{enumerate}[leftmargin=*]}
\newcommand{\ole}{\end{enumerate}}
\newcommand{\li}{\item}
\newcommand{\para}[1]{\paragraph{\textnormal{\textbf{#1}:}}}
\newcommand*{\rom}[1]{\expandafter\@slowromancap\romannumeral #1@}
\usepackage{romannum}

\begin{document}
% \let\WriteBookmarks\relax
% \def\floatpagepagefraction{1}
% \def\textpagefraction{.001}

% \ExplSyntaxOn
% $\keys_set:nn { stm / mktitle } { nologo }$
% \ExplSyntaxOff

\shorttitle{}

\shortauthors{Madhusudan Ghosh et~al.}

% \title [mode = title]{A Factored Transformer-based Sequence Labeling Model for Extracting Methodology Components from AI Research Papers}
%\tnotemark[1,2]

\title [mode = title]{Enhancing AI Research Paper Analysis: Methodology Component Extraction using Factored Transformer-based Sequence Modeling Approach}

\author[1]{Madhusudan Ghosh}[orcid=0000-0002-8330-2703]

%\author[1]{First Author\corref{cor1},~}
\cormark[1]


\ead{madhusuda.iacs@gmail.com}

\credit{Data curation, Formal analysis, Investigation, Methodology, Software, Validation, Visualization and Roles/Writing - original draft}

\affiliation[1]{organization={School of Mathematical and Computational Sciences, Indian Association for the Cultivation of Science },city={Kolkata},
postcode={700032},
country={India}
}

\author[2]{Debasis Ganguly}[orcid=0000-0003-0050-7138]
\ead{Debasis.Ganguly@glasgow.ac.uk}
\credit{Conceptualization, Formal analysis, Investigation, Methodology, Project administration, Supervision, Validation and Writing - review and editing}

\affiliation[2]{organization={School of Computing Science,  University of Glasgow},
city={Glasgow},
postcode={G12 8QQ},
    country={UK}}

\author[1]{Partha Basuchowdhuri}[orcid=0000-0001-7655-7591]
\ead{partha.basuchowdhuri@iacs.res.in}
\credit{Conceptualization, Formal analysis, Funding acquisition, Project administration, Resources, Supervision, Validation and Writing - review and editing}

\author[3]{Sudip Kumar Naskar}[orcid=0000-0003-1588-4665]
\ead{sudipkumar.naskar@jadavpuruniversity.in}

\credit{Conceptualization, Formal analysis, Project administration, Supervision, Validation and Writing - review and editing}

\affiliation[3]{organization={Department of Computer Science and Engineering, Jadavpur University},
    city={Kolkata},
    postcode={700032},
    country={India}}

\cortext[cor1]{Corresponding author}
%\cortext[cor2]{Principal corresponding author}

%
\begin{abstract}
Research in scientific disciplines evolves, often rapidly, over time with the emergence of novel methodologies and their associated terminologies. While methodologies themselves being conceptual in nature and rather difficult to automatically extract and characterise, in this paper, we seek to develop supervised models for automatic extraction of the names of the various constituents of a methodology, e.g., `R-CNN', `ELMo' etc. The main research challenge for this task is effectively modeling the contexts around these methodology component names in a few-shot or even a zero-shot setting. \hlt{The main contributions of this paper towards effectively identifying new evolving scientific methodology names are as follows: \romannum{1}) we propose a factored approach to sequence modeling, which leverages a broad-level category information of methodology domains, e.g., `NLP', `RL' etc.; \romannum{2}) to demonstrate the feasibility of our proposed approach of identifying methodology component names under a practical setting of fast evolving AI literature, we conduct experiments following a simulated chronological setup (newer methodologies not seen during the training process);
\romannum{3}) our experiments demonstrate that the factored approach outperforms state-of-the-art baselines by margins of up to 9.257\%
for the methodology extraction task with the few-shot setup.
} 

% More specifically, we show that our approach which learns to predict the sequence labels along with the category information achieves state-of-the-art performance on downstream methodology extraction task.
%
%and 8.756\%, in terms of recall and F-score.
%\romannum{3}) More precisely, we investigate whether the models trained on historical data (previously published papers) can then be incrementally retrained, with the predictions made by those models on new papers, such that it can predict more accurately in the future on newer papers.

\end{abstract}

\begin{highlights}
% \item \hlt{We investigate the feasibility  of incorporating category information of the scientific documents from the external Knowledge Base, which can introduce a significant performance gain on the transformer-based models, to identify the methodology names from the scientific literature as a sequence labeling task. To this end, we propose a novel transformer based factored model for sequence labeling task by partitioning the label space according to the category information.}

\item{ We propose a novel transformer-based factored model for our extraction task.}

%\item \hlt{As research in scientific discipline advances rapidly, occurrence of new methodology for any downstream task is inevitable. So, it will be a realistic and challenging task to understand how well a model, trained on the data points from a certain time period, performs on the data samples introduced after that particular time period. Thereby, we propose a novel chronological evaluation setup as opposed to widely applied conventional static setup of previous methodology extraction tasks from AI literature.}

\item{We propose a novel chronological evaluation setup proposed as opposed to static one.}

% \item \hlt{Building a new model from scratch every time, using newly introduced dataset, can be very much time consuming and resource intensive due to the fast advancements in scientific research. To address this issue, we propose an incremental retraining strategy to assess how effectively can already trained models be updated by including their own predictions as weakly supervised data for emerging methodology extraction from scientific documents.}

\item{We explore the viability of retraining technique for new methodology extraction task.}

% \item \hlt{Results of our experiments show that our proposed factored model achieves state-of-the-art performance in terms of F-score and Recall by a significant margin.}

\item{Experimental results show that our novel model achieves state-of-the-art performance.}

\end{highlights}

% Keywords

\maketitle

\section{Introduction}
\label{ss:intro}
\label{chap:intro}

A general trend observed in the scientific literature of any discipline is that it grows at a rapid rate embracing new theories, methodologies and their empirical validations or refuting. Some disciplines, such as the biological science or behavioural  science, often rely on a technique called meta-analysis that combines evidences from a number of studies \citep{metaanalysis}. The steep increase of scientific publications makes it difficult to manually conduct evidence synthesis. Moreover, such rapid advances in scientific methodologies~\footnote{Since `methodology' is more of an abstract concept, for the purpose of this paper, we consider the names of the more concrete \textit{methodology components} as our objects of interest, e.g., `BERT' (a transformer), `Adam' (an optimisation method) etc. From hereon, we use the word `methodology' to actually refer to the various methodology components encountered in AI research papers.} create difficulty for researchers to maintain a comprehensive and updated knowledge of the recent literature, which is critical for academic tasks, such as developing novel research ideas, selecting the correct baselines \citep{DBLP:conf/ecir/BediPBC22}, or peer-reviewing others' research. As attempts to provide automated support for these activities, previous research has applied automated information extraction (IE) approaches for identifying key properties from a scientific article, e.g., while \citep{hou-etal-2019-identification} proposed a supervised sequence labeling approach to identify task and dataset names from papers, \citep{Scells} and \citep{Singh:2017} applied an information retrieval based approach for analysis of systematic reviews.

\begin{figure}[t]%
    \centering
    \subfloat[\centering \small 2003 to 2017]{{\includegraphics[width=.4\columnwidth]{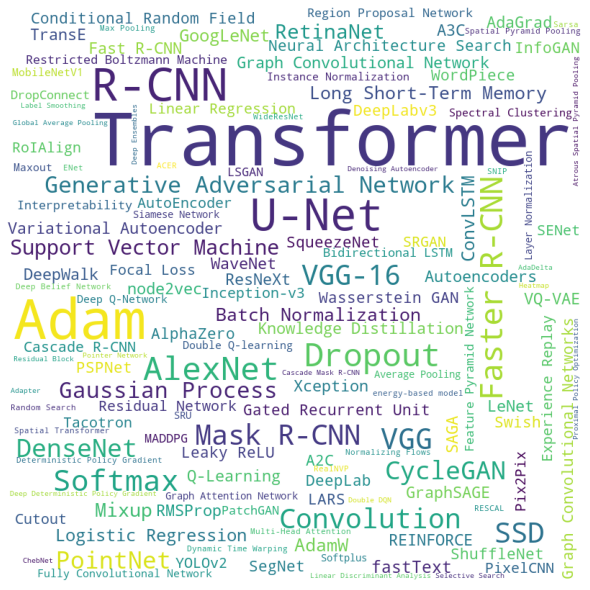} }}%
    %\qquad
    \subfloat[\centering \small 2018 to May, 2022]{\includegraphics[width=.4\columnwidth]{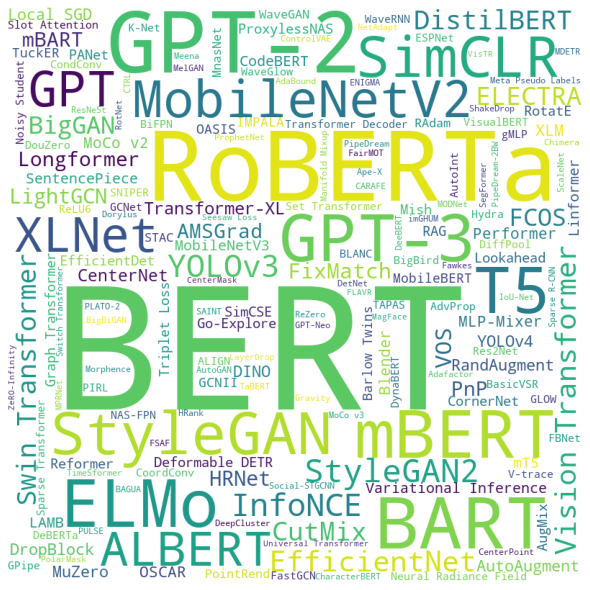} }
    \caption{\small Word cloud showing the relative change in the dominant methodology names between papers published in pre- and post-2017
    (data collected from the online repository of research articles - \url{paperswithcode.com}).
    %.\textcolor{cyan}{temporal significance of} evolving AI \st{methodologies} \emph{methodology components}. The \hl{charts} are prepared using the tags \textcolor{cyan}{and their frequencies during a time-span,} from the online repository of research articles - \url{paperswithcode.com}}.
    }
    \label{fig:indomain_outdomain}
\end{figure}

A limitation of the existing studies that seek to automatically extract key concepts, such as tasks, datasets and method names from scientific articles, is that either these have been empirically validated to work well only on the abstracts of papers \citep{gabor2017semeval}, or no attempt has been made to evaluate how effectively the models \emph{transition} to newer concepts and methods introduced into the literature.  
%{jain-etal-2020-scirex}
For instance, the context in which the word `transformer' is used in the recent years is substantially different from the conventional sense of the term; earlier it used to refer to an electrical device (Engineering) or an operator function (Mathematics), whereas these days it mostly refers to self-attention based neural architecture. Figure \ref{fig:indomain_outdomain} shows more examples from the NLP domain on how the dominant methodologies and their contexts can change over time.
%by the frequencies of occurrences of different methodologies in the NLP domain over time.

\begin{table}[t]
\centering
%\begin{adjustbox}{width=.99\columnwidth}
\small
%\scalebox{.75}{
\begin{tabularx}{\columnwidth}{@{}X@{}}
\toprule
\small Excerpt from paper \cite{vaswani2017attention} \\
\midrule
The \textbf{\underline{Transformer}} follows this overall architecture using stacked \underline{multi-head attention} and \underline{point-wise}, \underline{fully connected layers} for both the \underline{encoder} and \underline{decoder}... \\
%\cmidrule(r){2-2}
\midrule
\small Excerpt from paper \cite{devlin-etal-2019-bert} \\
\midrule
\underline{BERT}’s model architecture is a \underline{multi-layer bidirectional} \textbf{\underline{Transformer}} encoder based on the original implementation described in Vaswani et al. (2017) ...\\
\bottomrule
\end{tabularx}
%}
%\end{adjustbox}

\caption{
\small A demonstration on how the contexts around the methodology name - `transformer' evolved over time. During its inception, the word `transformer' needed to be explained in terms of stacked `multi-head attention', `point-wise fully connected layers' etc. After 2 years, with its frequent use in NLP and image classification, researchers simply refer to a `transformer' as a \textit{known concept}, and use it as a building block to develop novel architectures. This is the reason why the contexts around the word `transformer' in more recent papers, such as the BERT paper shown in the example, are different from the ones in 2017. These differences in contexts are likely to pose a challenge in automatically detecting the methodology name `transformer' from papers that appeared from around 2019 using a model that is trained on data annotated till 2017.
%Comparison between set of popular terms in different research domains till 2017 and after 2017.
}
\label{tab:Upto 2017 v/s After 2017}
\end{table}

While standard BERT-based~\citep{devlin-etal-2019-bert} sequence labeling approaches, e.g. \citep{beltagy-etal-2019-scibert,
jain-etal-2020-scirex} etc., when fine-tuned with the example mentions of the word `transformer' from scientific articles that appeared after the publication of the paper ``\textit{Attention is all you need}'' \citep{vaswani2017attention} in the year 2017 is expected to work fairly well to identify the neural network sense of the term, a critical pragmatic question to ask is: \textit{how well can we identify methodology names that are either not seen during the training process, or even if seen their meanings were different?} 
%In other words, how effectively can we identify `transformer' as a neural architecture component on a set of mention names from articles appearing before 2017?
%

This is an important question to ask, because an affirmative answer to this question implies that newer (domain-specific) methodologies of the future can then potentially be detected only by leveraging the information from the contexts of mentions of similar methodology names, e.g., the word `transformer' can still be detected as a \emph{methodology component} (corresponding to the neural network sense of the term) given that the contexts around its mentions are similar to the contexts of other similar methodology components existing before 2017 used to solve similar downstream tasks, e.g. LSTMs and CNNs. However, this is a challenging task because it is not only the introduction of new terminologies but also the evolution of common interpretations of terminologies that can pose further challenges. For instance, as Table \ref{tab:Upto 2017 v/s After 2017} illustrates how the context terms around the word `transformer' has changed since the inception of the methodology in 2017 from a formal definition to a more common usage.

Existing work has either focused on extracting a set of relatively well-defined entities, the contexts of which are less likely to evolve over time (e.g., tasks and datasets as investigated in~\citep{hou-etal-2019-identification,hou-etal-2021-tdmsci}), or they have investigated the more challenging task of methodology extraction from within a static collection~\citep{jain-etal-2020-scirex}. In contrast, we follow a more pragmatic course in our experiments where we investigate the feasibility of how effectively can methodology component names be identified under a zero or a few-shot setting.

% Moreover, different from the existing work on entity extraction from static collections of scientific papers, we investigate a more challenging problem - that of investigating the feasibility of weakly supervised approaches in extracting novel scientific entities. In contrast to existing studies, such as \cite{hou-etal-2019-identification,hou-etal-2021-tdmsci}, we focus on the methodology names. \textcolor{red}{This is more challenging because it is usually the case that newer methods are evaluated on well-defined benchmarks (combination of existing task and datasets), and also because of the fact that even new tasks and datasets are reported with the help of similar textual context, e.g., ``Our experiments are conducted on $\langle$DATASET\_NAME$\rangle$...'', or ``We evaluate our method on the $\langle$TASK\_NAME$\rangle$ task...'' etc.}\todo{Need to discuss}

% \hl{The problems to solve in this space are manifold}. Although a few distantly supervised datasets are available, none of these particularly cater to the problem that we address in this work.

%Those datasets do not contain the fine-grained detailing that our work demands.

\subsection{Practical Implications}
\label{ss:prac_impl}
\hlt{In this section, we highlight the practical implications of our research.
Due to the fast evolving nature of scientific discipline, it is difficult for researchers to manually keep a comprehensive track of the state-of-the-art knowledge of the scientific literature. Different task-specific tools have been proposed to mitigate this effort, which includes work in selecting correct baselines, peer-reviewing other research work~\citep{DBLP:conf/ecir/BediPBC22} and also scientific paper recommendation system~\citep{chaudhuri2022share}. 
For developing such types of automated systems, emerging methodology extraction from scientific articles plays a significant role, which is in fact, the objective of this paper}. 

\hlt{
Towards addressing the more challenging extraction task with the zero or few-shot setup, we, as a novel contribution in this paper, propose
a factored approach of sequence labeling by leveraging the broad level domain name categories of AI papers, e.g., NLP, RL (reinforcement learning) etc.
More concretely speaking, we partition either the input space constructing an ensemble of domain specific models, or we partition the label space allowing a multi-task oriented learning of predicting the text spans of relevant mentions along with the domain categories.
To tackle the task of practical interest, i.e., the downstream prediction task of learning from the past data and allowing the model to update itself from more recent data, we investigate how well the proposed factored models can be updated with incremental training on new silver-standard data - those accumulated from the model's predictions on recent scientific articles. To illustrate this with an example, consider training a model with gold-standard data till time $t$ (e.g., the year 2017), making predictions with this model on $t+1$ (e.g., the year 2018), feed back this predicted data of identified mentions as a silver-standard data to update the model, and then further use it on more recent data at time $t+2$ (e.g., the year 2019)~\citep{zeng2018large,qu2022distantly,luo2020appraisal}. Our experiments show that this strategy works 
particularly well for our proposed factored approaches.     
}

\subsection{Our Contributions}
To summarise, the following are our contributions in this paper.
\begin{enumerate}[leftmargin=*]

\item We propose two factored approaches to transformer-based sequence labeling - one that uses partitioning of the input space, and the other that uses partitioning of the label space. Both these models are shown to perform better than SciREX \citep{jain-etal-2020-scirex}, a state-of-the-art extractor of scientific concepts.

\item We propose our novel evaluation framework by partitioning the data chronologically. We carry out a more pragmatic approach of methodology extraction from AI papers, as opposed to the more conventional static setup of previous works \citep{hou-etal-2019-identification,hou-etal-2021-tdmsci,luan-etal-2017-scientific,jain-etal-2020-scirex}. In particular, we follow a chronological train-test split of the data, and investigate the effectiveness of various models in identifying new methodology names.

\end{enumerate}

The structure of the paper is as follows. Section~\ref{ss:rel_work} provides an overview of related work. Section \ref{ss:methodology} describes our proposed novel factored transformer-based approach for methodology name extraction.
%Our proposed novel chronological evaluation framework as opposed to the more conventional static setup is formalized in Section \ref{ss:retrain}.
In Section \ref{ss:eval}, we describe the experiment setup.
%the datasets and enumerate the research questions, followed by results and the experiment setup.
Section~\ref{ss:res_analysis} presents and analyses the observations from the results of our experiments.
Finally, Section~\ref{ss:conclsn} concludes the paper with directions for future work.

\section{Related Work}
\label{ss:rel_work}
Since our work mainly investigates deep metric learning based entity extraction approaches from scientific literature, we now first discuss the state-of-the-art techniques used for the named-entity recognition (NER) task, and then follow it up with how these NER models are adapted particularly for the scientific concept extraction task, a task which we, in fact, address in this paper.

\para{Generic NER} Our work revolves around solving the NER task, as we extract scientific entities, such as AI methodology names. Over the years, researchers have used many different techniques for entity extraction from different domains~\citep{li2020survey, nasar2021named, ji2020end, zhang2019whose}. Generally, NER problem is treated as a sequence labeling task~\citep{fei2019recognizing, ji2020end}. Popular deep learning based approaches
 such as LSTM based models~\citep{chiu2016named,cho2019biomedical} and CNN based models~\citep{sung2021cnnbif, ma-hovy-2016-end} and along with that attention module based frameworks~\citep{patra2019weakly, li2020attention, lin-etal-2020-triggerner, dutta2020changing} are widely used in literature to handle the sequence labeling task. Recently, neural network based approaches also incorporate probabilistic graph models such as Conditional Random Fields (CRF) in conjunction with BiLSTM as BiLSTM-CRF~\citep{mayhew2020robust}. Currently pretrained language model based approaches have become very popular. Some of the most important ones include intra-span and inter-span information~\citep{wang2020pre}, few-shot slot tagging~\citep{ma2021frustratingly}, decoupled NER model with two stage training~\citep{hu2021toward}, noise aware training mechanism~\citep{huang2021named}, contextual embedding model~\citep{hoory2021learning}, Trigger Matching Network which encodes and softly grounds the \emph{entity triggers} of unlabeled sentences \citep{lin-etal-2020-triggerner}, meta self-training framework for few-shot sequence labeling~\citep{wang2021meta}, and subsequence-based deep active learning~\citep{radmard2021subsequence}.
 Other extensions of NER
%There are some works that added new dimensions to the task of NER which
include biomedical NER \citep{tong2021multi,fan2020adverse,liu2021hybrid,asghari2022biner}, nested NER \citep{xu2021supervised,li2020recursively}, joint entity and relation extraction \citep{dai2019joint,zeng2020copymtl,nayak2020effective,xiao2020joint,sun2021progressive,li2021joint}, named entity normalization \citep{ji-etal-2021-neural}, unified NER \citep{li2021unified}, NER for low-resource languages \citep{DBLP:journals/talip/DasGG17}, document level NER using multitask learning approach~\citep{wang2021learning}, NER with multi-level topic aware attention mechanism~\citep{ma2023sequence}, information extraction module using multi-modal approach~\citep{toledo2019information}. Another approach is to align the entities by combining both embedding and symbol based techniques~\citep{jiang2022combining}. Additionally, an effective methodology, named FLAIR, was developed in~\citep{akbik-etal-2018-contextual} that allows users to fine-tune any word embedding and any PLM to yield improved results on the NER task.

%\citet{white2017inference, poliak-etal-2018-collecting} represented different NLP tasks, including extraction tasks, as natural language inference (NLI) \cite{bowman-etal-2015-large} problems. \citet{eichler2017generating, obamuyide2018zero} used NLI approaches for relation extraction. The best performing NLI model was introduced by \citet{liu-etal-2019-multi} for three-way classification, using BERT \cite{devlin-etal-2019-bert} with an accuracy of 91.9\%. \citet{kim2019semantic} used densly-connected co-attentive recurrent neural network to obtain 90\% accuracy.
%\\
%\citet{singh2019automated} performed search over publications and composed a leaderboard for a queried triplet, but they did not take advantage of textual content of the papers.
%\citet{milosevic2019framework, ghasemi2018tabvec,wei2006table,herzig-etal-2020-tapas} used \hl{tables for retrieving information}. \citet{pinto2003table} tried distinguish between captions, headers and rows in a stream of text.
%\\
%\citet{kardas-etal-2020-axcell} applied table type classification, table segmentation and linking results to leaderboards to automatically extract results from machine learning papers. \citet{hou-etal-2019-identification} extracted absolute metric values alongside the metric names, tasks and datasets. The authors used \hl{text experts} as well as direct tabular information to make inferences from table contents.
%\\\

%\\
      
%\\

\para{Weakly Supervised NER} In the present era of data-hungry deep learning models, availability of annotated data in adequate quantities is a problem. Preparing annotated data in a distantly supervised manner~\citep{mintz-etal-2009-distant,riedel2010modeling} alleviates this problem to some extent. In our work, we have generated weakly supervised silver-standard annotations for NER task to incrementally re-train our model. It prompted us to explore the state-of-the-art techniques for NER using weak supervision.
~\citet{mintz-etal-2009-distant} and ~\citet{riedel2010modeling} addressed the problem of relation extraction and mention identification without labelled text.
%Since then, many weakly-supervised approaches have been proposed for information extraction.
Recently, gazetteer based distant supervision technique had been applied to generate data for entity extraction task~\citep{mengge-etal-2020-coarse, yang2018distantly, cao-etal-2019-low,peng-etal-2019-distantly}. They used Wikipedia anchors or gazetteers and other existing knowledge bases (KBs) (i.e., MEDLINE, WikiData etc.) for this task.
\citet{ali2020fine} proposed FG-NET to obtain fine-grained named entity typing over distantly supervised data.
Furthermore, \citet{liu2020hamner} proposed a headword amplified multi-span distantly supervised method for NER. \citet{heist2021information} employed rule mining to populate Wikipedia-based KBs, where the rules were derived with a weak supervision approach.

\para{Emerging NER} For our scientific entity extraction task, we often come across newly coined terms (methodology names) or rarely used terms that eventually emerged as a frequently used term at a later time. Related works in emerging NER help us to understand how we could design our models to mitigate the challenge derived from the emerging nature of a number of entities.
\citet{allan1998line} conducted online identification of new topics, and also tracked how the meanings of these topics change over time. Recently, some important works have been done on the topic of language dynamics i.e., to how word meanings change over time~\citep{mitra-etal-2014-thats, mukherjee2011aging, maity2012opinion, loreto2012origin, endris2017dataset, singh2017structure, kelley2022using}. Recently, \cite{kabongo2023zero} addressed emerging TDM ($\langle$\texttt{Task}, \texttt{Dataset}, \texttt{Metric}$\rangle$) extraction task from the scholarly documents of AI literature by considering it as textual entailment recognition task. In contrast to identifying emerging tasks and datasets \citep{kabongo2023zero}, our work focuses on extracting newly introduced methodology components from AI literature, which is more challenging because emergence of new methodologies (models) is often more common than the emergence of new tasks and datasets. Moreover, in contrast to modeling this as a textual entailment task \citep{kabongo2023zero}, we propose a factored sequence labeling model.
%our work is even more challenging because the occurrence of new methodologies in literature is much more frequent than the occurrence of new TDM tuple.\todo{Need to discuss}

 % worked on language dynamics, i.e., to identify how word meanings change over time. %Recently, \citet{SocialDisNER} organized a shared task on identifying new disease names from tweets\footnote{\url{https://temu.bsc.es/socialdisner/}}. Although it successfully identified disease names, intuitively it was probably because of the high quantity of data manually labelled by experts.

% \cite{brambilla2017extracting} proposed a rule-based method to find new entities from social media text and incorporate them into existing KBs. 
%has been released namely SMM4H 2022 – Task 10.
%as released on \displaydate{date1}}

\para{Scientific NER} In recent years, IE from scientific papers has become a popular area of study in NLP. Our work deals with extraction of AI methodologies from scientific literature and led us to subsequent exploration of related works.
\citet{athar2012context, viswanathan2021citationie} used citations and \citet{jurgens2018measuring} used topic trends to extract information from scientific papers. Recently, \citet{safder2020deep} extracted algorithmic metadata from scholarly articles by applying Bi-LSTM. \citet{tsai2013concept} applied unsupervised bootstrapping method to identify and cluster the main concepts of a paper. In SemEval 2017 (Task 10), a novel key-phrase boundary classification method was presented to identify entities with class labels - process, materials and task~\citep{augenstein-sogaard-2017-multi}.

% introduced a new task, SemEval 2017 Task 10, the objective of which was to identify three types of (Keyphrase-like) entities -- \textit{Tasks}, \textit{Methods} and \textit{Materials}, and two relation types -- \textit{hyponym-of} and \textit{synonym-of}, from a corpus of 500 paragraphs from papers in the field of Computer Science, Material Sciences and Physics.

\citet{gabor2017semeval} proposed the task of semantic relation extraction and classification from scientific papers (SemEval 2018 Task 7) using a dataset of 350 annotated abstracts. 
These datasets had been used to develop their neural models for IE on scientific literature~\citep{ammar2017ai2, luan-etal-2017-scientific, augenstein-sogaard-2017-multi}.
\citet{luan-etal-2018-multi} used relation types and cross-sentence relations to construct a framework called Scientific Information Extractor (SciERC) for extraction of scientific entities of six types $\langle$\texttt{Task}, \texttt{Method}, \texttt{Metric}, \texttt{Material}, \texttt{Other-Scientific Term}, \texttt{Other-Generic}$\rangle$ and seven relation types.
\citet{jain-etal-2020-scirex} extended this work and released a dataset on document level IE which covers extraction of scientific entities of four types $\langle$\texttt{Dataset} , \texttt{Method}, \texttt{Task} , \texttt{Metric}$\rangle$.
\citet{hou-etal-2021-tdmsci} and \citet{kabongo2021automated} proposed a TDM tagger using a novel data augmentation technique. \citet{hou-etal-2021-tdmsci} applied this tagger to around 30,000 NLP papers from the ACL anthology and demonstrated its efficacy to construct an NLP TDM knowledge graph. Recently, an end to end large scientific knowledge graph has been been proposed by introducing `evaluatedby' and `evaluatedon' relationships~\citep{mondal2021end}. \hlt{Recently, \citet{ghosh2022astro} devised transfer learning strategy by finetuning the mT5 language model for downstream astronomical entity extraction task from astrophysics literature.} \citet{kardas-etal-2020-axcell} linked automatically extracted results from tables to leaderboards. Recently,~\citet{lo2020s2orc} and~\citet{saier2020unarxive} proposed two versions of large datasets namely `S2ORC', `unarXive' from different scientific disciplines. Additionally, citation context plays an important role for scientific information extraction task. \hlt{In this context, \citet{cohan2019structural} explored structural scaffolds based approach for citation intent classification task from scientific literature. \citet{lahiri2023citeprompt} used propmt based approach using GPT-2 for the detection of citation context from the scientific article.}

In addition to the extracting information from computer science articles as cited above, information extraction from articles of social science is comparatively under-explored. Works in this direction include those of \citet{shen2023sscibert} who apply a pre-trained language model specifically designed for social science texts called SsciBERT.
Additionally, scientific context generation is also an important area of research along with the scientific IE task. Recently, \citet{chen2021scixgen} explored the contributions of context in scientific text generation by proposing two primary tasks, i.e., context-aware description generation and context-aware paragraph generation. \hlt{Due to fast evolving nature of scientific literature, another interesting area, which remains under-explored, is scientific paper recommendations. In relation to this context, \citet{chaudhuri2022share} proposed a novel scientific paper recommendation system. 
}
%to automatically extract results from machine learning papers. 

The main difference between our work and the existing studies of scientific entity extraction from research papers is that we study this problem in a realistic few-shot setup, where the objective is to identify new and emerging methodologies without a significant presence of such entities in historical data. Moreover, we propose a factored sequential modeling approach for the task which is also novel.
% We chronologically split our distantly supervised \PWC~dataset to carry out zero-shot future method components extraction from scientific literature so that evolving methodological terminologies can be tracked.
% Additionally, we make use of 7 topics from the \PWC~KB, (e.g., NLP, CV, etc.) which boosts our model performance.

%\citet{cao2021generativere} incorporate a generative model for joint entity and relation extraction, \citet{cabot2021rebel} present an autoregressive approach for relation extraction, ,   

\section{Proposed Methodology}
\label{ss:methodology}
%\todo{What's not clear to me is how do you know which sections are what? This seems to be a challenging problem in itself.}

%Following previous work 
% The main task of our work is that, given an AI related scientific paper of a particular year, we are interested in extracting the new intricated emerging terminologies related to methodology from the sentences of that particular document. 
% Although scientific papers are generally very lengthy documents, only a few parts of the document play a vital role in identifying methodological entities. Hence we represent a document in \textbf{DocAIMER} representation.

% \paragraph{DocAIMER.} Each scientific paper's \textbf{DocAIMER} representation comprises four essential sections: Abstract, Introduction, Methodology, ExpSetup and Results. Abstract, Introduction, and Methodology help to identify the proposed neural method and its associated parts. Similarly, ExpSetup and Results also help to extract the terminologies related to the experimental setup, i.e., optimizer, regularization, loss function, etc.

\begin{figure*}[t]%
    \centering
    %\vspace{.6cm}
    \subfloat[\centering \small \Factored]{{\includegraphics[width=.5\textwidth]{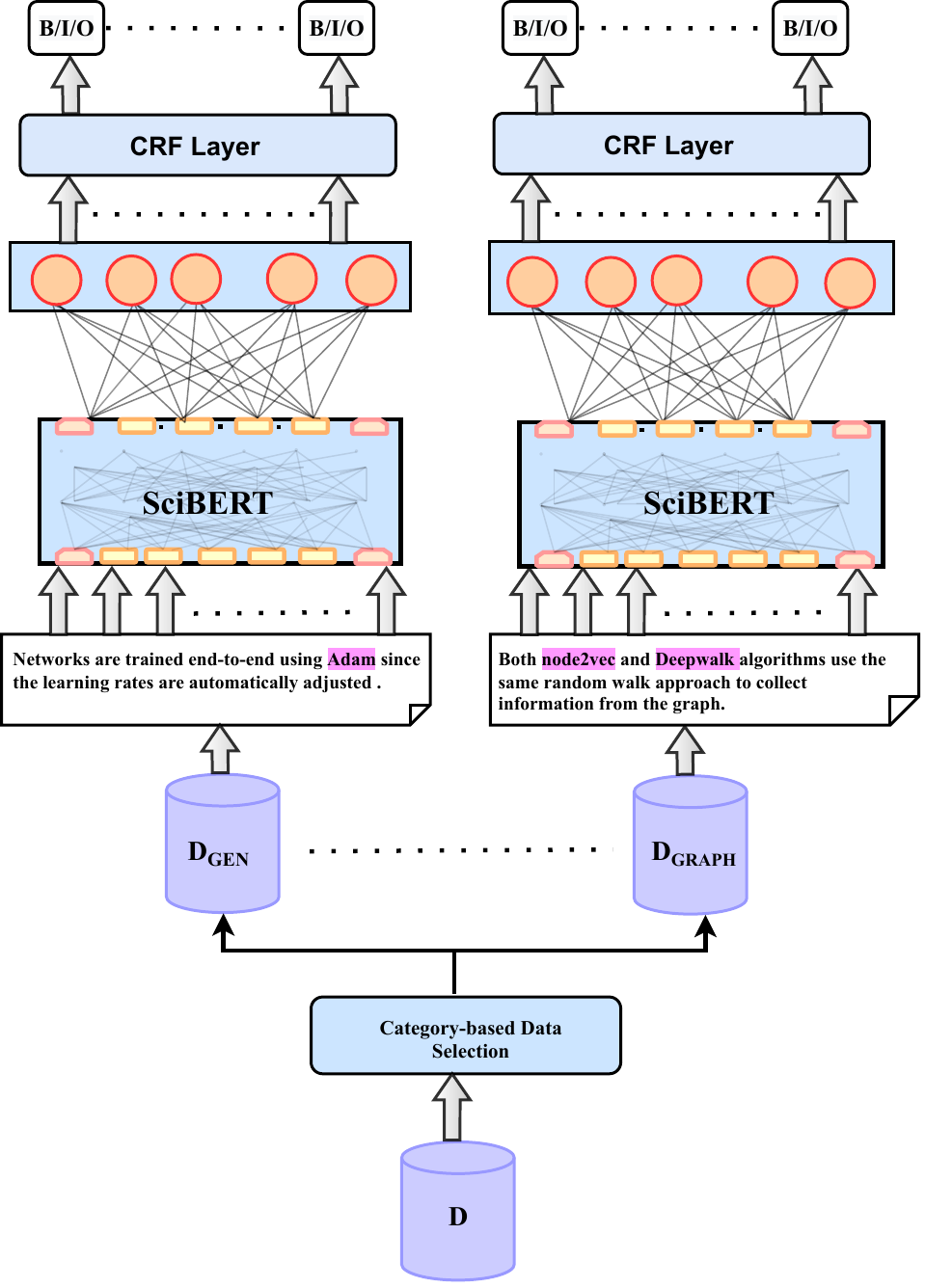} }
    \label{fig:CG-model}
    }%
    %\qquad
    \subfloat[\centering \small \FG]{{\includegraphics[width=.5\textwidth]{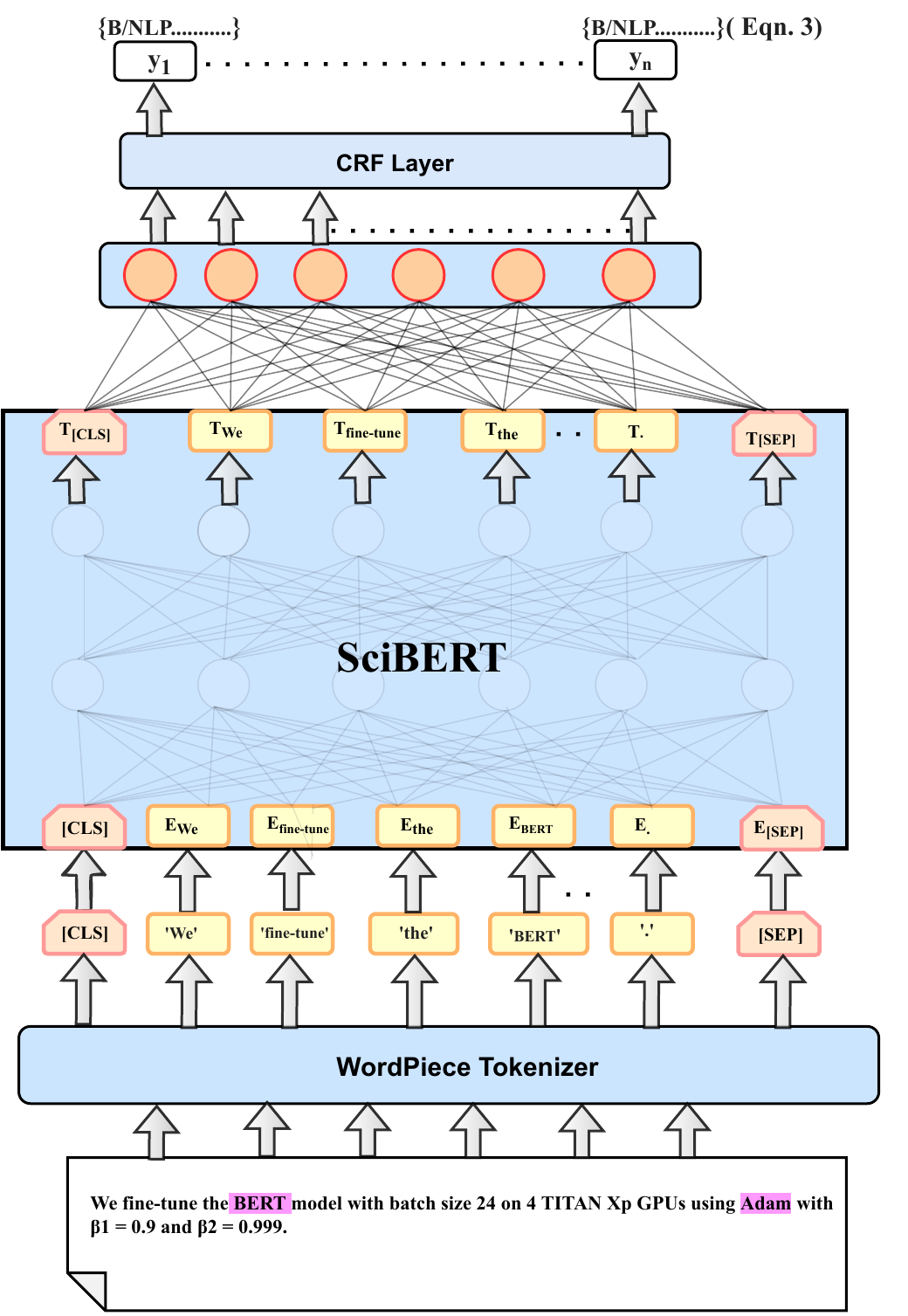} }
    \label{fig:fac-CG}
    }%
    \caption{\small Schematic of the two factored models proposed for methodology extraction - \textbf{\textit{a})} input partitioning \textbf{\textit{b})} label partitioning.}
    \label{fig:cg_model}
\end{figure*}

% Screenshot of \PWC~webpages showing \textbf{a)} the broad categories or the topics of a paper, and also its \textbf{b)} methodology component tag names as uploaded by the authors. \textbf{c)} Per-category distribution of methodology name tags across a chronological split of the data (year = 2017).
%In this section, we describe the details of our proposed neural architecture.
This section starts with the problem definition considering the methodology extraction as a sequence labeling task, which is then followed by details on the neural architecture of our proposed factored models.

\para{Sequence Labeling}
Given a sentence from
%\st{the DocAIMER representation of}
a document, we follow a sequence labeling approach for method component extraction. Formally, given a sentence (word sequence) $x=(\mathrm{w_1,w_2...w_n})$,
the objective is to learn a function parameterized by $\theta$, $f_{\theta}$, that maps an observed sequence of embedded vectors to a sequence of labels
%\begin{equation}
%\label{problem_definition}
%\begin{aligned}
$f_{\theta}:\vec{w_1},\ldots,\vec{w_n} \rightarrow y_1,\ldots,y_n$,
%\end{aligned}
%\end{equation}
where each $\vec{w_i} \in \mathbb{R}^d$ is an embedded vector of the token $w_i$, and each $y_i \in \{\mathrm{B}, \mathrm{I}, \mathrm{O}\}$ denotes a label which, as per the standard NER terminology, indicates if it is the beginning, continuation or the end of a text span (in the context of our work - an identified methodology name).
Given a set of examples of such $\mathcal{D} = \{(x, y)\}$ sequence pairs, the parameters $\theta$ of a sequence classification model are learned by optimizing
\begin{equation}
z = \underset{\theta}{\mathrm{argmin}}\, \sum_{(x, y)\in \mathcal{D}}\mathcal{L}(y,f(x,\theta)),
\label{eq:problem_definition_wloss}
\end{equation}
where $\mathcal{L}$ is a standard loss function, e.g., the cross-entropy.

% \begin{equation}
%     \mathcal{L(\theta)} = -\frac{1}{N}\sum_{i=1}^{N}\mathrm{y_i.log\hat{y_i}}
%     \label{loss_cg}
% \end{equation}s

% Where $y_i$'s are the actual labels and $\hat{y_i}$ are the predicted labels from our model. $N$ is the total number of training instances. As we are using standard \textbf{BIO} tagging scheme to detect the entity span we name this model as \textbf{C}oarse-\textbf{G}rained model (\textbf{CG}). Our model \textbf{CG} model is depicted in Figure. \ref{fig:CG-model}.  So, given a sentence instance from the $\mathcal{D}$\textsubscript{indomain}, we prepare standard input representation using the WordPiece module and pass it into the SciBERT module to fine-tune the SciBERT language model for our downstream sequence labeling task. During inference time, we just freeze the pre-trained model weights and note the performance metric score against $\mathcal{D}$\textsubscript{outdomain} instances. 

% $\theta$ are model parameters which are to be learned during end-to-end training of our neural model. 
 %of Equation \ref{eq:problem_definition_wloss}.

We employ the state-of-the-art BERT-based transformer architecture which leverages the information obtained during the pre-training phase, transferring this knowledge to facilitate a downstream task \cite{wang2020adaptive}. In particular, following \citet{jain-etal-2020-scirex}, we use SciBERT \citep{beltagy-etal-2019-scibert} as our pre-trained transformer model and fine-tune it on the sequence labeling objective of Equation \ref{eq:problem_definition_wloss}.

% \todo{Don't describe the dataset prep here.}
% To identify intricated new emerging methodological terminologies from sentences of any scientific articles, we apply different natural language processing techniques and also seek to devise a transfer learning strategy on a pre-trained language model.

\begin{comment}
To incorporate the temporal essence into the task of extracting new methodological terminologies from the recent scientific articles, we split our whole data corpus into two parts based on a specific year. We call these two datasets in-domain data and out-domain data. 
During dataset construction, we also prepare a mapping of methodological terms and their corresponding year from the \PWC~knowledge base (KB) (i.e., we take that year of a method name whenever it first time appears in PWC KB). We are using that mapping to create our in-domain and out-domain datasets. We first split the method components for a given year into two parts. If a method appears before or in that particular given year, we keep it in one part and vice versa. Then we apply some search heuristics using those two sets of method components on our distantly supervised dataset to prepare our in and out-domain sentence instances accordingly.
\end{comment}

\subsection{Factored Model for Sequence Labeling} \label{ss:factored-models}
%\hlt{
We propose two factored modelling approaches for the task of methodology names extraction by leveraging their broad-level categorical information (more details in Section \ref{ss:method_cat}).
Both the proposed models involve the common processing steps of fine-tuning a transformer model and applying a standard CRF decoder layer similar to \cite{jain-etal-2020-scirex}.
%To develop our proposed neural frameworks, we consider fine-tuning strategy on \textcolor{magenta}{a} BERT based model followed by a standard CRF decoder layer (e.g., as in \cite{jain-etal-2020-scirex}) in an end-to-end fashion as our basic model. We introduce our factor approaches as an extension of our basic model.}
%
% \st{The \PWC~KB provides an additional information on the broad category of each method name (cf. Figure).
% Fine-tuning a BERT based model with the inclusion of a CRF decoder layer is unable to leverage this information. To alleviate this, in our work, we propose two different extensions of the basic model that is able to leverage this information.}
%
Both these models make use of partitions in the data induced by the category information; while one approach partitions the input data, the other partitions the label space. More details follow.

\subsubsection{Model with Partitioned Input Space}
\label{ss:factored-models-i/p}

Let there be a total of $m$ broad-level categories that may be associated with the methodology names, e.g., `GPT' for `NLP', `ResNet' for `computer vision' etc. (Section \ref{ss:method_cat} describes how these categories are obtained for the PapersWithCode dataset). In this factored approach, we partition each sentence of the dataset into $m$ categories, and then train $m$ different sequence labeling models for each of these partitions. Formally speaking, 
\begin{equation}
f_{\theta_i}(x\in \mathcal{D}_{i})\mapsto y\in \{\mathrm{B},\mathrm{I},\mathrm{O}\},\,\,\,i=1,\ldots,m,\,\,\text{and}\,\,
\mathcal{D}=\bigcup\limits_{i=1}^{m}\mathcal{D}_{i}.
\label{eq:factored_i/p}
\end{equation}
In other words, we $m$ different models $f_{\theta_i},\ldots,f_{\theta_m}$ models, where each $f_{\theta_i}$
takes as input instances of the $i^{\text{th}}$ category and tunes the corresponding $\theta_i$ as required model parameters using the sequence labeling loss as defined in Equation~\ref{eq:problem_definition_wloss}.

The common intuition behind applying this approach is that a single model may not essentially capture the inherent diversities of the contexts of different types of methodology names across different domains. For instance, the context of the methodology component `transformer' is expected to be different for image and text categories - the former expected to be associated with contexts relevant to image/video classification, whereas the latter being associated with terminologies such as summarization, question answering, etc. As described later in Section \ref{ss:method_cat}, the value of the category level information is available at the level of documents (research articles). The document-level category is then assigned to each of its constituent sentences during the model training phase.
%}

After training $m$ different models, during inference phase we assume that the category of a test document is known. This allows determining which model, out of the $m$ choices, is to be used for prediction.
We call this \textbf{DFG} based approach
%\hlt{/\textbf{\FLRFactored}}
(as presented in Figure~\ref{fig:CG-model})
to indicate that it uses
a partition of the input \textbf{D}ata to model the
\textbf{F}ine-\textbf{G}rained information.
\begin{figure}[t]
\centering
%\vspace{-1cm}
\includegraphics[width=0.99\columnwidth]{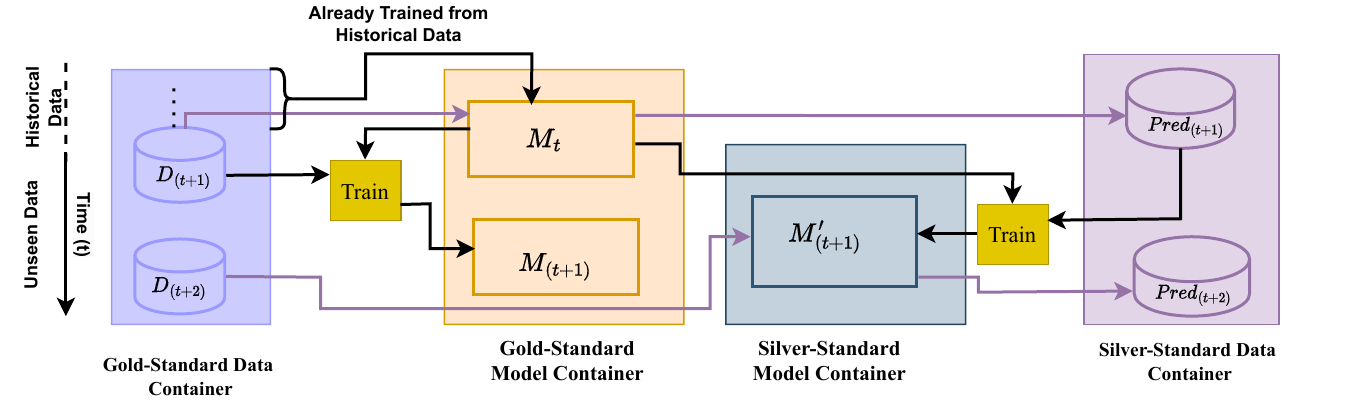}
\caption{\small Schematic diagram of retraining strategies on chronological data partitions.}
\label{fig:retraining_pic}
\end{figure}

\subsubsection{Model with Partitioned Label Space}
\label{ss:factored-models-o/p}

This model instead of using the three coarse-grained labels $\{\mathrm{B},\mathrm{I},\mathrm{O}\}$ (indicators of the beginning, continuation and end of a text span), uses the category information to create more fine-grained labels. Specifically speaking, if $c$ denotes one of the $m$ categories, for each coarse-grained indicator label (B/I/O), we create its fine-grained counterpart. Formally, a label's value $y$ belongs to the following set:
\begin{equation}
\{\langle Z,c\rangle:\,\,Z\in \{\mathrm{B},\mathrm{I},\mathrm{O}\}\land c\in\{1,\ldots,m\}\}.
\label{eq:fg-label-defn}
\end{equation}
To give an example, if `NLP' represents a category, then the B/I/O labels corresponding to this category are $\langle B, \mathrm{NLP}\rangle$ etc.
We then use the labels as defined in Equation \ref{eq:fg-label-defn} to train the sequence labeling objective.
We name this approach \textbf{LFG} to indicate that it uses a set of \textbf{L}abels that are \textbf{F}ine-\textbf{G}rained (see Figure~\ref{fig:fac-CG}). 
Note that in contrast to the DFG model, the LFG one involves training only a single model overall, which means that it is not required to know the category information during the inference phase. Since the downstream task involves identifying the text segments only, the fine-grained labels post-inference are trivially mapped back to their coarse-grained interpretations, i.e., $\langle Z,c\rangle \mapsto Z$, e.g., $\langle \text{B}, \mathrm{NLP}\rangle$ and $\langle \text{B}, \mathrm{GRAPH}\rangle$ are treated as identical by stripping off the categories `NLP' and `GRAPH' after the inference step. 
%For instance, a text span `Adam' identified with the predicted fine-grained tags - $\langle B, \mathrm{NLP}\rangle$, $\langle I, \mathrm{NLP}\rangle$ $\langle O, \mathrm{NLP}\rangle$ is equivalent to another text segment `Adam' from a different sentence identified with the predicted fine-grained tags - $\langle B, \mathrm{GRAPH}\rangle$, $\langle I, \mathrm{GRAPH}\rangle$ $\langle O, \mathrm{GRAPH}\rangle$
%
%\hlt{
%For instance, if our LFG based model predicts `Adam' as a methodology component with some category then during the post inference phase we simply drop the category information and treats it as a detected methodology component.
%}

% In this section we investigate different neural architectures for the entity recognition encoder $f_{\theta}$, all fine-tuning experiments are built on top the of the widely used SciBERT~\cite{beltagy-etal-2019-scibert} pre-trained language model. Throughout all experiments, we consider in-domain sentence instances as training data and out-domain instances as test data. In this work we have written entity recognition task and sequence labeling task interchangeably. 

%\input{EMNLP 2022/figdefs/fg_model_architecture}

% \paragraph{Domain-specific training.}

%\section{Parameter Settings}
%\label{experimental_settings}
\section{Chronological Evaluation Framework}
\label{ss:retrain}
In this section, we propose a framework to evaluate various methodology extraction models in a way different from the conventional one-time evaluation using a train:test split of the data \cite{hou-etal-2019-identification,jain-etal-2020-scirex}. The main motivation of proposing a new evaluation framework is to investigate the practical use-case on how effectively can models trained on historical data can adapt themselves to keep pace with the fast evolving literature of a scientific discipline. This is particularly important because obtaining annotated data for gold-standard model training takes considerable manual effort \cite{liang2020bond,zhao2021glara}. From a pragmatic point-of-view, a scientific concept extraction model should thus be robust enough to be trained incrementally with its own predicted entities (silver-standard aka weakly supervised training) over a period of time. 

More concretely speaking, the core idea of the proposed evaluation framework is to first induce a chronological split of the dataset comprising scientific articles. The first (earlier in time) part of the split is treated as the existing literature and a sequence labeling model $M$ to identify scientific concepts is trained on this gold-standard data (denoted by $M_t$ in Figure \ref{fig:retraining_pic}). Next, $M_t$ is applied on the documents published in the year $t+1$ (denoted as $D_{t+1}$ in Figure \ref{fig:retraining_pic}) to extract a set of predicted scientific concept names $Pred_{t+1}$. This data is then fed back to $M_t$ as silver-standard data, and $M_t$ is incrementally retrained with these automatically extracted entities to yield the next version of the model, $M'_{t+1}$~\cite{luo2020appraisal}. Subsequently, $M'_{t+1}$ is applied on the next year's data $D_{t+2}$ and the cycle continues.

An expected observation from an effective model is that retraining it with silver-standard data should be better than not updating the model, e.g., $V(M'_{t+1}, D_{t+2}) > V(M_t, D_{t+2})$, where the notation $V(M, D)$ denotes some evaluation metric (e.g., precision, recall etc.) obtained with model $M$ on dataset $D$. However, training with silver-standard data for a subsequent time-period (i.e., in our case with a model's own predictions) is also expected to be worse than the ideal situation of the availability of annotated (gold-standard data) over this period. In other words, following our terminology, it is expected that $V(M'_{t+1}, D_{t+2}) < V(M_{t+1}, D_{t+2})$.
Later in our experiments in Section \ref{ss:incremental}, we show that our proposed models indeed exhibit this behaviour, i.e., they may be updated with silver-standard data (its own predictions) to improve its effectiveness on future research papers.

In particular, the chronological split of the data to simulate the delineation of the data into the past history and the future, we choose the year 2017, i.e., papers dated up to 2017 (inclusive) serving as the training split, and papers that are published from 2018 to date serving as the evaluation or test data. The specific reason for choosing $t=2017$ is to capture the general effect of the transition in several domains around the year 2017 after the introduction of transformers, explainable AI etc. For example, the NLP community itself has witnessed a paradigm shift from the pre-BERT era (2018) to the recent times.

The decision to assign a paper to one of the splits was determined by its timestamp. The timestamp of a paper is the most recent timestamp of its constituent methodology names. Thus, if there exists at least one methodology name in a paper the timestamp of which is past 2017, it is allocated to the evaluation set. For example, although a paper may have used LSTMs (a pre-2018 technology) to model a sequence of BERT (a post-2017 technology) vectors of passages, the paper is still allocated to the evaluation set.

Figure \ref{fig:percat-stats} shows the number of methodology name tags per-category across the training and the test data splits. It can be seen that the prediction task is likely to be challenging because the number of unique tags appearing post-2017 is often higher than the number of tags appearing pre-2018.

% \section{Dataset Preparation}
% \label{ss:dataset}
% \input{Sections/dataset}

\section{Experiment Setup}
\label{ss:eval}
% \begin{figure*}[t]%
%     \vspace{-2cm}
%     \centering
%     \subfloat[\centering PwC KB Method Components Page]{{\includegraphics[scale=.1,keepaspectratio=true]{EMNLP 2022/Figure/PwC_method_merged_page.png} }
%     \label{fig:pwc_page_1}
%     }%
%     %\qquad
%     \subfloat[\centering PwC KB Method Components Paper Page]{{\includegraphics[scale=.35,keepaspectratio=true]{EMNLP 2022/Figure/PwC_Page_merge} }
%     \label{fig:pwc_page_2}
%     }%
%     \caption{PwC Page Example}
%     \label{fig:pwc_page}
% \end{figure*}

\begin{figure*}[t]%
    % \vspace{-2cm}
    \centering
    \subfloat[
    %\centering
    %PwC KB Method Components Page
    \small Categories
    ]{{\includegraphics[width=.48\columnwidth]{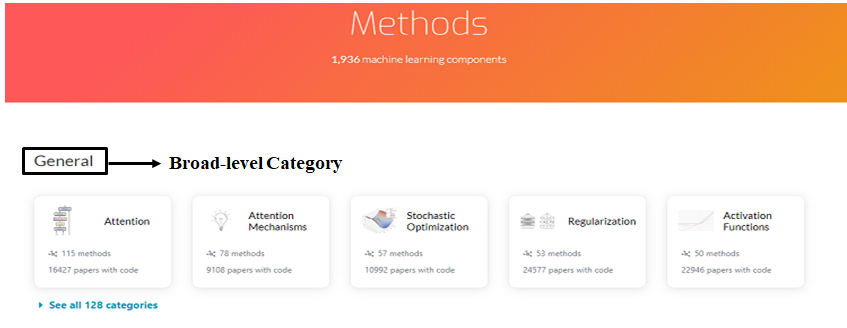} }
    \label{fig:pwc_page_1}
    }%
    \subfloat[
    \centering
    %PwC KB Method Components Paper Page
    \small Methodology tags of a sample paper.
    ]
    {{\includegraphics[width=.48\columnwidth]{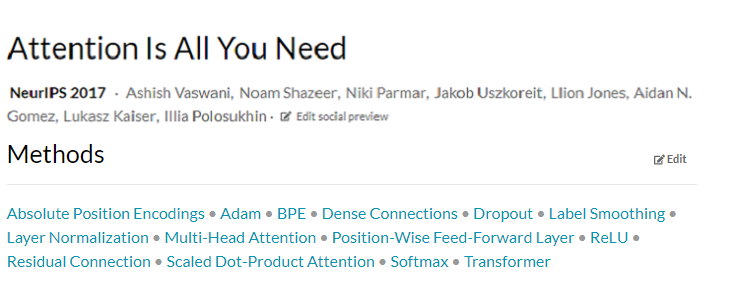} }
    \label{fig:pwc_page_2}
    }%
    % \subfloat[
    % \centering
    % %PwC KB Method Components Paper Page
    % \small Per-category statistics.
    % ]
    % {{\includegraphics[width=.33\columnwidth]{Figure/data_fig_2.pdf} }
    % \label{fig:percat-stats}
    % }%
    \caption{\small Screenshot of a \PWC~webpage showing \textit{a)} the broad categories or the topics of a paper, and also its \textit{b)} methodology component tag names as uploaded by the authors.}
    \label{fig:pwc_page}
\end{figure*}

 % \textit{c)} Per-category distribution of methodology name tags across a chronological split of the data (year = 2017).

% \begin{figure}[t]%
%     % \vspace{-2cm}
%     \centering
%     \subfloat[
%     %\centering
%     %PwC KB Method Components Page
%     \small Categories
%     ]{{\includegraphics[width=.7\columnwidth]{Figure/pwc_new.png} }
%     \label{fig:pwc_page_1}
%     }%
%     \\
%     \subfloat[
%     \centering
%     %PwC KB Method Components Paper Page
%     \small Methodology tags of a sample paper.
%     ]
%     {{\includegraphics[width=.55\columnwidth]{Figure/PwC_Page_merge} }
%     \label{fig:pwc_page_2}
%     }%
%     \\
%     \subfloat[
%     \centering
%     %PwC KB Method Components Paper Page
%     \small Per-category statistics.
%     ]
%     {{\includegraphics[width=.65\columnwidth]{Figure/data_fig_2.pdf} }
%     \label{fig:percat-stats}
%     }%
%     \caption{\small a) Screenshot of a \PWC~webpage showing \textit{a)} the broad categories or the topics of a paper, and also its \textit{b)} methodology component tag names as uploaded by the authors. \textit{c)} Per-category distribution of methodology name tags across a chronological split of the data (year = 2017).}
%     \label{fig:pwc_page}
% \end{figure}

In this section, we first describe the details of preparing the dataset used in our experiments from the PapersWithCode (PwC) collection, and then we follow it up with the research questions pertaining to the task of methodology extraction, and the methods investigated towards addressing those questions.

\subsection{Dataset Preparation}
\label{ss:dataset}
% In this section, we describe the steps towards constructing the dataset for our experiments. 
% \hlt{ Our main goal is to extract emerging methodologies from AI literature. Therefore, it is important to create a dataset in chronological fashion from the scientific articles. We start this section with a description of how the dataset is prepared from the \PWC~corpus. We then describe how the documents are represented and the weak annotations are constructed from the \PWC~tagged methodology components.}

We employ a realistic setup for evaluating our proposed models in a `few-shot' manner. Specifically, we employ a chronological splitting strategy of the entire dataset to constitute a training set of older technologies and a newer (more recent) evaluation set comprised mostly of new methodology names that are non-existent in the training set (zero-shot setup). In situations where the methodology name phrase exists in the training set, it is mostly the case that the semantic meaning of the string has evolved significantly, e.g., the word `transformer'.
%Due to the scarcity of annotated datasets in scientific disciplines, to train our novel factored-based model and evaluate it in our proposed novel chronological set up.
This section outlines the dataset preparation strategy and provides detailed statistics on the prepared dataset.
% \begin{figure}[!t]
% \centering
% \includegraphics[width=.65\columnwidth]{EMNLP 2022/Figure/data_fig_1.png}
% \caption{}
% \label{fig:data_1}
% \end{figure}

% \begin{figure}[t]%
%     %\vspace{-2cm}
%     \centering
%     \subfloat[
%     \centering
%     \#Unique tags
%     ]{{\includegraphics[width=.49\columnwidth
%     %scale=.1,keepaspectratio=true
%     ]{EMNLP 2022/Figure/data_fig_2_log.pdf} }
%     \label{fig:fig_data_1}
%     }%
%     %\caption{No. Unique Method Components Per Category}
%     %\qquad
%     \subfloat[\centering \#Unique tags in the two chronological splits]{{
%     \includegraphics[
%     width=.49\columnwidth
%     %scale=.35,keepaspectratio=true
%     ]{EMNLP 2022/Figure/data_fig_1_log.pdf} }
%     \label{fig:fig_data_2}
%     }%
%     \caption{Per-category distribution of methodology name tags in the \PWC~dataset.}
%     \label{fig:pwc_page}
% \end{figure}

\begin{figure}[t]
    \centering
    \includegraphics[width=.55\columnwidth]{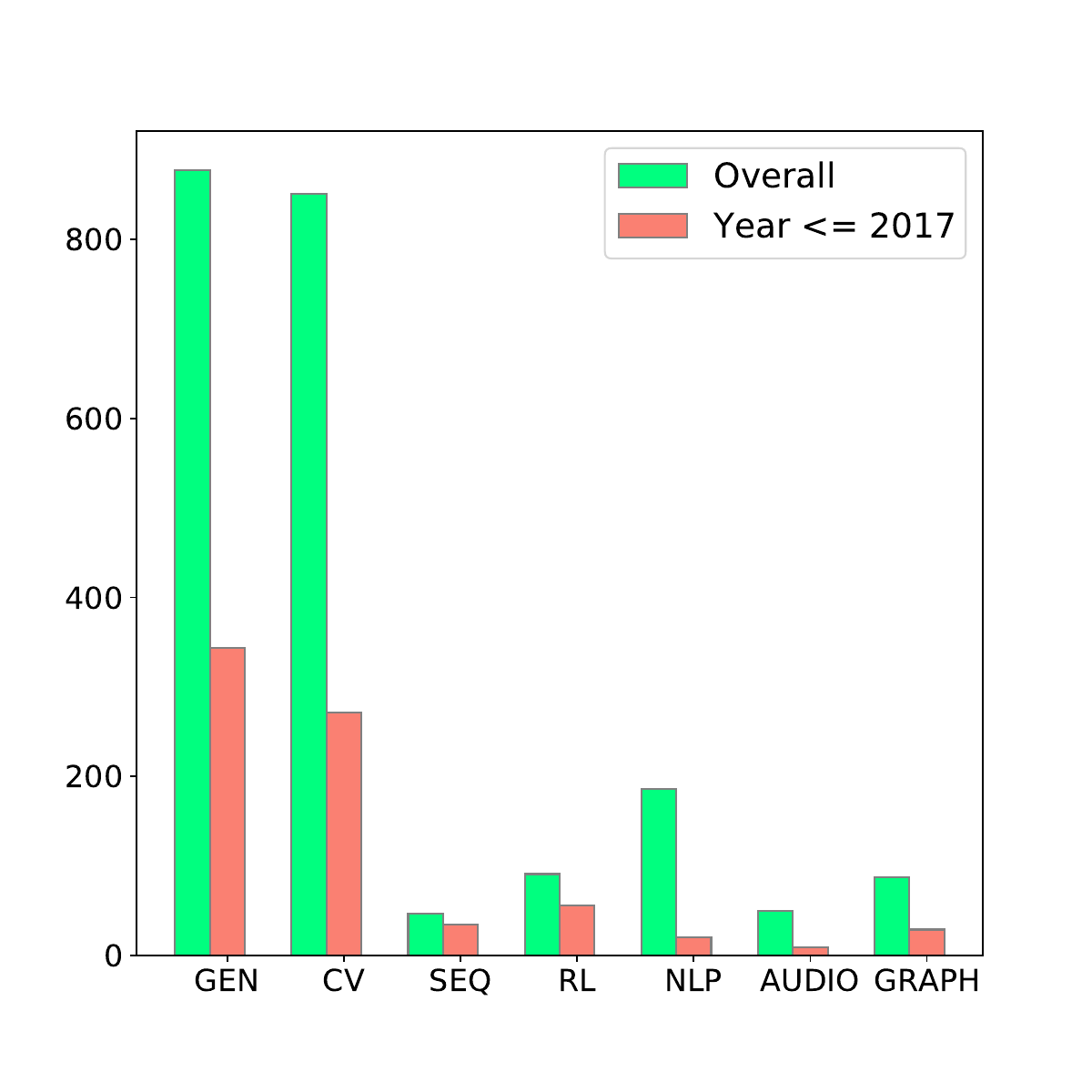}
    \caption{\small Per-category distribution of methodology name tags across a chronological split of the data (year = 2017).}
    \label{fig:percat-stats}
\end{figure}

%.695

% \begin{table*}[!t]
% % \hspace{-0.5cm}
% \begin{tabular}{|c|c|c|c|c|c|c|c||c|}
% \hline
% Categories & GEN & CV & SEQ & RL & NLP & AUDIO & GRAPH & Total \\
%   \hline \hline
%  Total PwC tags & 1,403 & 1,438 & 100 & 150 & 296 & 92 & 184 & 3,663 \\ 
%   \hline
%   \# of tags found in text & 877 & 851 & 47 & 91 & 186 & 50 & 87 & 2,189 \\
%  \hline \hline
%  \end{tabular}
%  \caption{Tags collected from PwC (after case normalization) vs. exact matches for those tags in the text extracted from the PDFs.}
%  \label{tab:methodCompStatindomain_stat}
%  \end{table*}

% \begin{table}[]
% \begin{tabular}{lllllllll}
% \hline
% \textbf{Category}        & \textbf{Gen} & CV   & \textbf{SEQ} & \textbf{RL} & \textbf{NLP} & \textbf{Audio} & \textbf{Graph} & \textbf{Total} \\ \hline
% Total PwC tags           & 1403         & 1438 & 100          & 150         & 296          & 92             & 184            & 3663           \\ 
% \# of tags found in text & 877          & 851  & 47           & 91          & 186          & 50             & 87             & 2189           \\
% \hline
% \end{tabular}
% \caption{Tags found from PwC (after case normalization) vs. exact matches for those tags in the text extracted from the PDFs.}
% \end{table}

\begin{table}[t]
\centering
\small
\begin{tabular}{c|c|c}
\hline
{} & {\bf Std. dev}  & {\bf Mean} \\
\hline
The distribution of papers over time &3709.4 &2152.1\\
Distribution of number of method tags per paper & 4.6 &6.7\\
Distribution of distinct tags present in the dataset & 510.7  & 98.7\\
\hline
\end{tabular}

\caption{\small Detailed statistics of the dataset used in our experiments.}
\label{tab:Dataset1}
\end{table}
%\para{Selecting and parsing Papers from PwC}
\subsubsection{Selecting and parsing Papers from PwC}
As a first step, we downloaded the \PWC~JSON dump\footnote{\url{https://production-media.paperswithcode.com/about/papers-with-abstracts.json.gz} as updated on \displaydate{date}} comprised of a total of 291,503 papers. Each paper is accompanied with metadata that includes the title, abstract, a list of tags indicating methodology components such as `RoBERTa' and `GPT-2' etc., and the Arxiv links to the paper. Since the information most important to us is the mentions of the methodology components for a paper (as these correspond to the $y$ values used to train our sequence labeling models as per Equations \ref{eq:problem_definition_wloss} to \ref{eq:fg-label-defn}),
%later on used for obtaining the weak annotations as described later in Section \ref{ss:weakannot}),
we remove those papers which do not possess this metadata field. This leaves us with a total of 34,560 papers to work with. Table \ref{tab:Dataset1} presents some statistics of the dataset. 
As an example, Figure \ref{fig:pwc_page_2} shows the set of methodology names tagged by the authors from a sample \PWC~page.

%
%\paragraph{Using SciPDF Parser on the PDF text.}
As the next step, we use the SciPDF parser\footnote{\url{https://github.com/titipata/scipdf_parser.git}}, a parser implementation tailored for scientific PDFs, to parse the text from each PDF file. The parser outputs a structured view of the document text separated out into segments comprising sections, figures and tables. In contrast to the previous work of \cite{hou-etal-2019-identification}, we exclude the captions of figures and tables because it is unlikely that a methodology name will only appear in a figure or a table caption without occurring within the text of a document.

%\para{Method Categories}
\subsubsection{Method Categories}
\label{ss:method_cat}
% Sentences were extracted from the text repository using a simple regex code so that our models could be provided with sentence-level input. NLTK tokenizer was used to identify the tokens from sentences. 
% A total of 1,884 unique method component names were found from PwC.
The broad-level method categories that our factored models use (Equations \ref{eq:factored_i/p} and \ref{eq:fg-label-defn}) is obtained from the PwC metadata. In particular,
%\hlt{
each document in the PwC corpus is associated with a broad-level category or topic (in our case $m=7$ as mentioned in Section \ref{ss:factored-models-i/p} and \ref{ss:factored-models-o/p}), which is one of General (GEN), Computer Vision (CV), Sequence-to-Sequence Modeling (SEQ), Reinforcement Learning (RL), NLP, Audio \& Speech (AUDIO), and Graph-based Modeling (GRAPH). These document-level tag values are assigned to each constituent sentence of a document.
Figure \ref{fig:pwc_page_1} shows an example of methodology instances from the `General' category of the \PWC~KB.

\subsubsection{Document Representation and Weak Labeling for Training} \label{ss:weakannot}

%\para{Document Representation}
%\hlt{
We first segment an article into different sections, such as `Introduction', `Experiments', etc. The text from some of these sections is more likely to contain the entities that are sought to be extracted. Prior work used the content from four different sections of a paper, namely the title, abstract, experiments, and table data to identify tasks, datasets and evaluation measures and their values \cite{hou-etal-2019-identification}; they named their document representation as DocTAET: \underline{T}itle, \underline{A}bstract, \underline{E}xpSetup, and \underline{T}ableInfo.
However, different from \cite{hou-etal-2019-identification}, our task is to extract the methodology components from a paper; we, therefore, represent a paper as concatenated chunks of text from the following sections: `\underline{A}bstract', `\underline{I}ntroduction', `\underline{M}ethodology', `\underline{E}xperiments' and `\underline{R}esults', and use their initials as per the convention of \citet{hou-etal-2019-identification} to name the representation as DocAIMER. Note that different from DocTAET, in DocAIMER we exclude the TableInfo field and include the content from the Introduction; this is because the method names are more likely to be present in the abstract, introduction, methodology, the experiment and the result sections of a paper, and are not likely to occur exclusively in the captions (without these concepts being mentioned in the text).

%}
%+++DG: Don't have that stats yet
% In fact, this is what we observed in the occurrences of the method names in the text of the articles. \todo{Avg. frequency of mention names in the respective sections will be good.}

Different from DocTAET, we do not include data from tables because our intention is not to extract the evaluation metrics or scores. Instead, different from the DocTAET representation, we include the `Introduction' section since it is likely that it develops the motivation for a paper's methodology using references to other method names in the literature. Following the argument of \cite{jain-etal-2020-scirex}, we also leave out the `Related Works' since it primarily describes earlier published works rather than the current paper's work. Similarly, we include the experiment setup section because it is likely to be comprised of the more fine-grained architectural details of a proposed methodology, e.g., information on the optimizer, regularization, loss function, etc.

To identify the textual content of each paper section, we use a simple rule-based lookup to exclude sections that are not likely to be one of the AIMER-type sections, e.g., we exclude sections with names `Background', `Related work', `Conclusions', `Conclusions and Future work' etc. For reproducibility, we release the dataset for research purposes.

 \begin{table}[t]
%\vspace{-.5cm}
\centering
\small
%\begin{adjustbox}{width=.9\columnwidth}

%\begin{tabular}{c|ccc|ccc}
\begin{tabular}{@{}l@{~~~}l ccc| ccc@{}}
\toprule
\multirow{2}{*}{} & \multirow{2}{*}{Model}
& \multicolumn{3}{c}{90:10 split on $\leq$2017}             & \multicolumn{3}{c}{Train: $\leq$ 2017, Test: $>$ 2017}
\\
%\cmidrule(lr){1-7}
\cmidrule(r){3-5}
\cmidrule(r){6-8}
% \cmidrule(r){9-11}
% \cmidrule(r){12-14}
& & Prec.      & Recall      & F-score     & Prec.     & Recall     & F-score \\
\midrule

% \multirow{8}{*}{Baselines} &
% $a$ &NQC \cite{kurland_tois12} & 
% 0.331 & 0.298 & 0.285 & 0.227 &
% 0.239 & 0.185 & 0.183 & 0.107 &
% 0.259 & 0.243 & 0.179 & 0.124 \\

% & $b$ & Clarity \cite{croft_qpp_sigir02} & 
% 0.173 & 0.248 & 0.172 & 0.207 &
% 0.156 & 0.147 & 0.096 & 0.113 &
% 0.239 & 0.215 & 0.107 & 0.129 \\

\multirow{4}{*}{\rotatebox{90}{Baselines}}
&\blstm & 0.7281          & 0.5018          & 0.5941          & 0.6095         & 0.1960          & 0.2966\\
&\blstmglove & 0.8257          & 0.5666          & 0.6720          & \textbf{\underline{0.6460}}         & 0.1933          & 0.2976\\
&\CG & 0.9632          & 0.9863          & 0.9746          & 0.3722         & 0.2380          & 0.2903
\\
 &\FLRCG & \textbf{\underline{0.9832}}          & 0.9881          &0.9857         &0.4383         & \textbf{0.2452}   &  \textbf{0.3145}\\
 &SciREX & 0.9831          & \textbf{0.9887}  & \textbf{\underline{0.9859}}         & 0.4604         &0.2177          &0.2956\\
\midrule
%\midrule

\multirow{4}{*}{\rotatebox{90}{Ours}}

&\FLRFactored  & \textbf{0.9821} & \textbf{\underline {0.9894}} & \textbf{0.9856} & \textbf{0.5789} & 0.1907         &0.2869\\
&\FLRCGBinary & 0.9790          &0.9836          &0.9813          &0.5784         &0.2365         &0.3357\\
&\FLRFG  &0.9821          &0.9887          & 0.9854         &0.5110         &0.2236         &0.3111\\

&\FLRFGBinary & 0.9816          &0.9891          &0.9853         &0.4742         &\textbf{\underline{0.2690}} & \textbf{\underline{ 0.3433}}\\

%\midrule

\multirow{4}{*}{\rotatebox{90}{Ablations}} &\Factored  & 0.9658 & 0.9886 & 0.9771 & 0.4930 & 0.1418         & 0.2203 \\
&\CGBinary & 0.9646          & 0.9754          & 0.9700          & 0.4363         & 0.1065         & 0.1712\\
&\FG  & 0.9616          & 0.9845          & 0.9729          & 0.4429         & 0.1938         & 0.2696   \\
&\FGBinary & 0.9628          & 0.9859          & 0.9742          & 0.4271         & 0.2590 &0.3225 \\
\bottomrule
\end{tabular}
%\end{adjustbox}
\caption{\small
%\hlt{
A comparison between the models investigated for the methodology extraction task. Evaluated conducted with two different train-test splits - (i) conventional 90:10 split, and (ii) chronological splitting corresponding to a few-shot setup. In few-shot setup, the improvements of the best results obtained with \FLRFGBinary (bold-faced underlined) vs. the best performing baseline, \FLRCG(bold-faced), are significant  (t-test with 95\% confidence).}
\label{tab:outdomain_stat}
\end{table}
\subsection{Research Questions}
\label{ss:rqs}

% We have described the problem we deal with in Section~\ref{chap:intro}. In accordance to that, we investigate the following research questions.
%\paragraph{Research Questions.}

%\input{tabledefs/example3}

%\uls
%\item
As discussed in Section~\ref{ss:intro}, the rapid growth of scientific disciplines has led to a frequent introduction of new methodologies in the literature. In context of this, we investigate our first research question,

\uls
\li \textbf{RQ-1}:How effectively can methodology extraction be conducted under the challenging settings of few-shot data for training (unseen or infrequently seen methodology names during training), and evolving contexts of the mentions?
\ule
We compare the performance of various models (baselines, our proposed and their ablations; see Section \ref{ss:proposed_model}) to answer this question.

% \noindent \textbf{RQ-1}: How effectively can we develop a model to predict \hlt{novel (previously unseen) and emerging (previously seen) \emph{methodology components}} \st{from a} \textcolor{magenta}{where the} training dataset \st{that }contains a few mentions \hlt{(few-shot setting)} of some entities (e.g. ``transformer'') or no mention \hlt{(zero-shot setting)} of some entities (e.g. ``BERT'')?\\

% \textbf{Special case of RQ-1}: \hlt{We have observed that in some cases \PWC~has missed out on marking relevant entities with \emph{methodology components}. As a consequence, such entities, when appearing in the test data, lead to a zero-shot setting. Can our model accurately identify such entities? We expect our model to be able to achieve such\todo{incomplete} }
% , with little or no presence    The most important question of this research work is to understand whether we can perform zero-shot extraction of novel unknown terms from emerging AI methodologies on the basis of previously known AI method components and relate them to a specific domain of research.
    
%\item
% In the second research question, we reiterate the motivation behind the approach explained in subsection~\ref{ss:factored-models}.
Our second research question is directed towards finding answer to the question on what variant of the factored approach - input data partitioned or label partitioned (Figure \ref{fig:cg_model}), works particularly well for the methodology extraction task. In particular,
\uls
\li \textbf{RQ-2}: What is the most effective way of leveraging the domain or category of a scientific article for training a sequence labeling model in a factorised manner?
\ule

In the next research question, we aim to understand whether including silver-standard data may help to improve a model (see Section \ref{ss:retrain}).
\uls
\li \textbf{RQ-3}: How effectively can models be updated with their own predictions included as silver-standard data for the task of methodology extraction from scientific documents?
\ule

%This is more of an analysis...
% In the final research question, we introspect about the evolution of the contextual meaning of the texts around the same entity over time, as shown in Table~\ref{tab:Upto 2017 v/s After 2017}. 

% \noindent \textbf{RQ-4}: \hlt{How accurately can our proposed model capture the shift of surrounding context (as explained in Table~\ref{tab:Upto 2017 v/s After 2017}) in relation to any methodology name used in AI literature?}

% \noindent \textbf{RQ-5}: \hlt{In training data entities have been marked with \emph{methodology component tags} by \PWC~(using human annotators), whereas in the upcoming scientific literature the entities are still not marked by default. }

% information Does domain-specific training It is also important to understand whether category based knowledge of AI method components helps in improving such a zero-shot extraction of novel terms from emerging AI methodologies. For example, learning the context of existing NLP related method components may help in extraction of novel unknown terms from NLP research domain.
    
% \item Another important question would be to enquire whether to use the category based knowledge of AI method components in a single model using labelled data, where the labels denote the categories as mentioned in Eq. \ref{eq:fg-label-defn}, or whether to split the input data into category based AI method components and related sentences to train different models for identifying emerging terms of different categories.

%\ule

\subsection{Methods Investigated}
In this section, we describe the details of each method investigated in our experiments.

%
% We perform successful zero-shot extraction of novel AI related terms as shown in Table~\ref{tab:outdomain_stat}. We carry out our experiments by splitting the \PWC~corpus using the year 2017 as the point of split. The reason behind our choice lies in the exponential growth of deep learning based methodologies around 2017. As a result the growth in number of papers in AI research domain has experienced a surge as seen in the \PWC~KB. Therefore, if we select the earlier years, as a point of split the number of papers and AI method components will be limited and insufficient to train good predictive models. \textcolor{red}{We do not choose recent years as a point of split because, in that case, the number of papers and related novel AI terms in the test data will be insufficient to understand the generalization capability of the predictive models.}
%
\subsubsection{Baselines}
In order to evaluate the effectiveness of our proposed novel factored-based models, we compare our methodologies with the following baseline methods. The baselines we have chosen are well-established and widely used in this field, providing a solid point of reference for our experiments~\citep{beltagy-etal-2019-scibert, schweter2020flert, jain-etal-2020-scirex}.

\uls
\li \textbf{\blstm}:
 Since we approach our methodology extraction task as a sequence labeling task (as discussed in Section~\ref{ss:methodology}), we employ a BiLSTM-based approach as it is considered to a standard technique for supervised information extraction widely used in the literature \citep{cho2019biomedical}. In particular, we employ a standard BiLSTM-based approach that operates with word embedded vectors as input. Specifically, for word embedding we employed GloVe~\citep{pennington-etal-2014-glove} with the dimensionality set to 100.
 
\li \textbf{\blstmglove~\citep{panchendrarajan2018bidirectional}}: 
This a variant of \blstm~where an additional CRF decoder layer is applied for the methodology extraction task with the other settings of \blstm~remaining identical. It has been shown in the literature that CRF used in combination with BiLSTM produces better results for sequence labeling tasks, such as NER~\citep{zeng2017lstm}.
%Here, we also employed 100 dimensional GloVe vector representation as our required word embedding.

\li \textbf{\CG~\citep{beltagy-etal-2019-scibert}}: \CG~is a widely used pretrained language model trained on scientific documents collections, and has been shown to yield state-of-the-art performance in information extraction tasks for the scientific domain \cite{beltagy-etal-2019-scibert}. For our experiments, we fine-tune the standard \CG~model on our labeled data (Section \ref{ss:weakannot}) with the sequence labeling loss (Equation \ref{eq:problem_definition_wloss}).
%Therefore, we adopt \CG~as the core language model and fine-tune it on our methodology extraction task using sequence labeling loss.

\li \textbf{\FLRCG~\citep{schweter2020flert}}: 
%\hlt{
Similar to augmenting \blstm~with CRF, in this approach we augment \CG~with a CRF layer.

\li \textbf{\scrx~\citep{jain-etal-2020-scirex}}: 
This approach, as proposed in \cite{jain-etal-2020-scirex}, augments the \CG~model with a layer of BiLSTM and CRF. The overall model is trained in an
end-to-end manner with the sequence labeling loss of Equation~\ref{eq:problem_definition_wloss}.

\ule

%\hlt{
The difference of our proposed approach with the baseline methods enumerated above is that none of them use the category label information towards addressing the methodology extraction task. This means that better results obtained with our methods can hence be attributed to this missing factor.
%To the best of our knowledge, we are the first to introduce factor based models using category label information from \PWC~KB. Futhermore, in our model we have used chronologically partitioned data as training and test set. Traditionally, these state-of-the-art methods do not use such a setup for training and testing purpose.
%}

\subsubsection{Our Proposed Methodologies}
\label{ss:proposed_model}
For all our proposed factor approaches, we employ the CRF layer in the model (the ones without the CRF layer are presented as ablations). We also employ different levels of granularity for a thorough investigation of the factored models; details follow.
% \hlt{We explore our factoring approach on transformer based model against our methodology extraction task by introducing the category information of the methodology components (as discussed in Section~\ref{ss:method_cat}). We propose two techniques to apply factoring approaches on transformer based architecture - one is partitioning the whole data samples into different categories (as defined in Equation~\ref{eq:factored_i/p}) and \st{another} \textcolor{magenta}{the other} one is partitioning the label space (as defined in Equation~\ref{eq:fg-label-defn}).
% In our case, we select \FLRCG~as our core model to conduct the factored experiments \st{because} \textcolor{magenta}{since} we observed in our initial experiments that \FLRCG~performs relatively \st{well}\textcolor{magenta}{better}. More details follow next.}

\uls
\li \textbf{\FLRFactored}:
%\hlt{
As per Section~\ref{ss:factored-models-i/p}, we fine-tune $m=7$ of different $f_{\theta_i}$s (Equation~\ref{eq:factored_i/p}) using 7 individual categories (Section~\ref{ss:method_cat}) of data instances. This approach routes a test instance sentence to the appropriate model and hence the assumption is that the category information is known during inference time (in our experiment setting, this is available from the PwC metadata).
%}

\li \textbf{\FLRCGBinary}: %\hlt{
This approach is similar to \FLRFactored, except that it uses the category information in a more coarse-grained manner. To be more precise, we merge together the 6 non-generic categories into one category. This means that along with the `GEN' category, we work with $m=2$ categories (hence the suffix B to denote binary). This approach thus follow a middle course between \FLRCG~and \FLRFactored~in terms of the granularity of the category information usage. This actually leads to a more balanced partitioning of the data input space as can be seen from
Figure~\ref{fig:percat-stats} that the `GEN' category alone comprises a large proportion of the data instances.
%A common intuition behind this from Figure~\ref{fig:percat-stats} shows that `GEN' and `CV' categories contain large number of instances. To handle the data imbalance problem from \textcolor{magenta}{the} input data partitioning perspective, we investigate this approach.%}

\li \textbf{\FLRFG}: %\hlt{
Different from \FLRFactored~(which involves input partitioning), this approach is the fine-grained version of the label partitioning model as discussed in Section~\ref{ss:factored-models-o/p}. Similar to \FLRFactored~, this uses more $m=7$ fine-grained labels, one for each category (Equation \ref{eq:fg-label-defn}) to fine-tune a sequence labeling model in an end-to-end manner.
%}

\li \textbf{\FLRFGBinary}~\footnote{Our source code and dataset are available at \url{https://github.com/Madhu000/ie-nlplit.git}}: %\hlt{
Again, similar to \FLRCGBinary, this takes a middle path between \FLRCG~and \FLRFG~in the sense that $m=2$ types of labels are used (GEN vs. the rest merged into one). %indicating the start and end of entities belonging to a particular category type.
Similar to \FLRCGBinary, this method also takes a more balanced course towards addressing the task specific to the PwC dataset, which is dominated by the GEN category.
%issue prevalent in our dataset as Figure~\ref{fig:percat-stats} shows that the `GEN' and `CV' categories have more number of datapoints than the other categories combined.%}
\ule

%\subsubsection{Ablations of our proposed Methodologies}
%\label{ss:ablations}
%\hlt{
For each proposed method outlined above we derive its corresponding ablated version by removing the CRF layer. We name each approach by removing the `CRF' subscript from its name, e.g., \FGBinary~instead of \FLRFGBinary.
\para{Implementation Details}
%\hlt{

Implementation of the baselines and the proposed approaches uses the FLAIR framework\footnote{\url{https://github.com/flairNLP/flair}}.
All the models were trained on Google Colab PRO plus.
In terms of the common neural network settings, we used AdamW~\citep{loshchilov2018decoupled} as the optimizer with a learning rate of $5e^{-5}$ and  a stopping criterion as mentioned in~\citet{conneau-etal-2020-unsupervised}; the training batch size used was 32. 
%}

% In all the CRF experiments, we utilize the FLAIR framework and train all the neural models using AdamW~\cite{loshchilov2018decoupled} optimizer with a very small initial learning rate of $5e^{-5}$ and a stopping criterion as mentioned in \cite{conneau-etal-2020-unsupervised}}. 
%During evaluation time we fix our batch size to 64.
%We used Google Colab PRO plus to carry out the experiments.

% Further, in Table \ref{tab:predicted-vs-actual} we present two sample contexts where the \FGBinary~model (the best performing one as seen from Table \ref{tab:Upto 2017 v/s After 2017}) makes a  
% \input{EMNLP 2022/tabledefs/example3}

\renewcommand{\para}[1]{\paragraph{\textnormal{\textbf{#1}}.}}
\section{Results}
\label{ss:res_analysis}

In this section, we present a comprehensive analysis of our research questions as mentioned in Section~\ref{ss:rqs}. We mainly focus on providing detailed observations and insights for our research questions.

\subsection{Model Comparisons}  \label{ss:summary}

 %\hlt{
 
%This set of experiments and corresponding results pertains to \st{our }\textbf{RQ-1} as raised in Section~\ref{ss:eval}. To investigate this, we train the baselines and our proposed models on the historical data (pre-2018 papers) and evaluate the models on papers from 2018 onward.
%In this setting, we evaluate the \st{effectiveness} \textcolor{magenta}{model performance} only on methodology entities that \st{never} \textcolor{magenta}{do not} appear\st{ed} in the training set and the entities \st{those} \textcolor{magenta}{that are} seen in the training set only for a few times. This is to ensure that these entities could not have been detected by a simple lookup operation. In addition to this, we also report results with a conventional setting of 90:10 train-test split on the pre-2018 dataset.

\para{Main results}
In relation to RQ-1, Table~\ref{tab:outdomain_stat} presents a comparison between the models investigated (best results in each group, i.e., the baselines or the novel models, which also includes the ablations, both bold-faced; additionally the overall best results across the two groups are underlined). It can be observed from Table~\ref{tab:outdomain_stat} that all baseline models along with our proposed factored extensions work satisfactorily well with the standard percentage-based train:test split over the dataset of papers till 2017 (which is an easier setup and widely reported in the literature, e.g., \cite{hou-etal-2019-identification,beltagy-etal-2019-scibert,hou-etal-2021-tdmsci,jain-etal-2020-scirex}).
The baseline models mostly works better than the proposed factored versions when it comes to the standard train:test (in-domain) evaluation.

However, under the challenging situation of a more realistic out-domain setup (chronologically partitioned data), it is observed that the factored versions of the models (our proposed ones) work better (this is seen from the fact that the recall and the F-score values in the lower group are underlined).
As a general observation, it is also seen that the factored models lead to better recall values.
An additional observation is that the CRF-based models work consistently better than their non-CRF counterparts.

% \begin{figure}[t]
% \centering
% \vspace{-1cm}
% \includegraphics[width=0.9\columnwidth]{Figure/result_fig_1.pdf}
% \caption{\small Per-category F-score values for each model on the chronological split of the data.}
% \label{fig:result_analysis}
% \end{figure}

% \begin{figure}[t]%
%     \centering \small \FLRCG~variants
    %\vspace{.6cm}
    % \subfloat[\centering \small \CG~variants]{{\includegraphics[width=.51\textwidth]{Figure/result_fig_2.pdf} }
    % \label{fig:result_analysis_1}
    % }%
    %\qquad
%    \includegraphics[width=.51\textwidth]{Figure/result_fig_new.pdf}
%     %\label{fig:result_analysis_2}
%     \caption{\small \hlt{Per-category F-score values for each of our proposed model along with the well performed baseline model on the chronological split of the data.}}
%     \label{fig:result_analysis}
% \end{figure}

%\small \Factored
%\small \FG

\begin{figure}[t]
\centering
\vspace{-1cm}
\includegraphics[width=0.6\columnwidth]{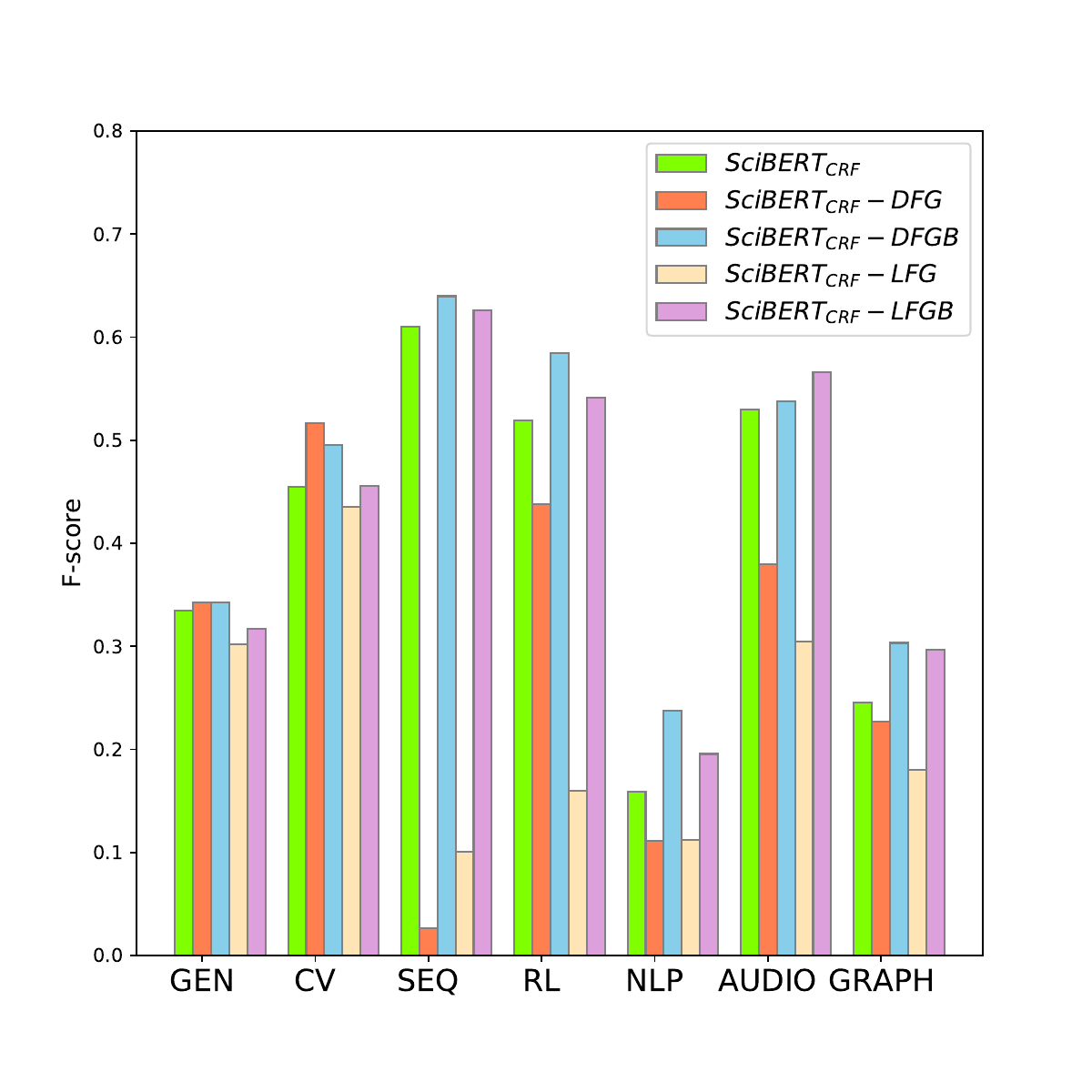}
\caption{\small
%\hlt{
Per-category F-score values obtained with the models investigated on the chronological split of the data.
%}
}
\label{fig:result_analysis}
\end{figure}

\para{Granularity of the factored models}
Turning our attention to RQ-2 (which explores what levels of granularity in the factored models work most effectively), we first observe from Table \ref{tab:outdomain_stat} that the binary version of the label-based factoring (\FLRFGBinary) performs the best for the few-shot (chronological split) setup, which means that a balanced factoring works particularly well for this more realistic experiment setup. 
Further analysis presented in Figure~\ref{fig:percat-stats} shows that this effectiveness measure of \FLRFGBinary~is actually computed over a substantial proportion of new methodologies encountered during the inference phase (zero-shot setup). In fact, for the SEQ and RL categories, this proportion is even higher than $0.75$.

\begin{figure}[t]%
    \centering
    %\vspace{.6cm}
    \subfloat[\centering \small
    %\hlt{
    %Evaluation including entities till 2017.
    %Performance on pre-2018 and post-2017 methods
    %}
    ]
    %{
    {\includegraphics[width=.45\textwidth]{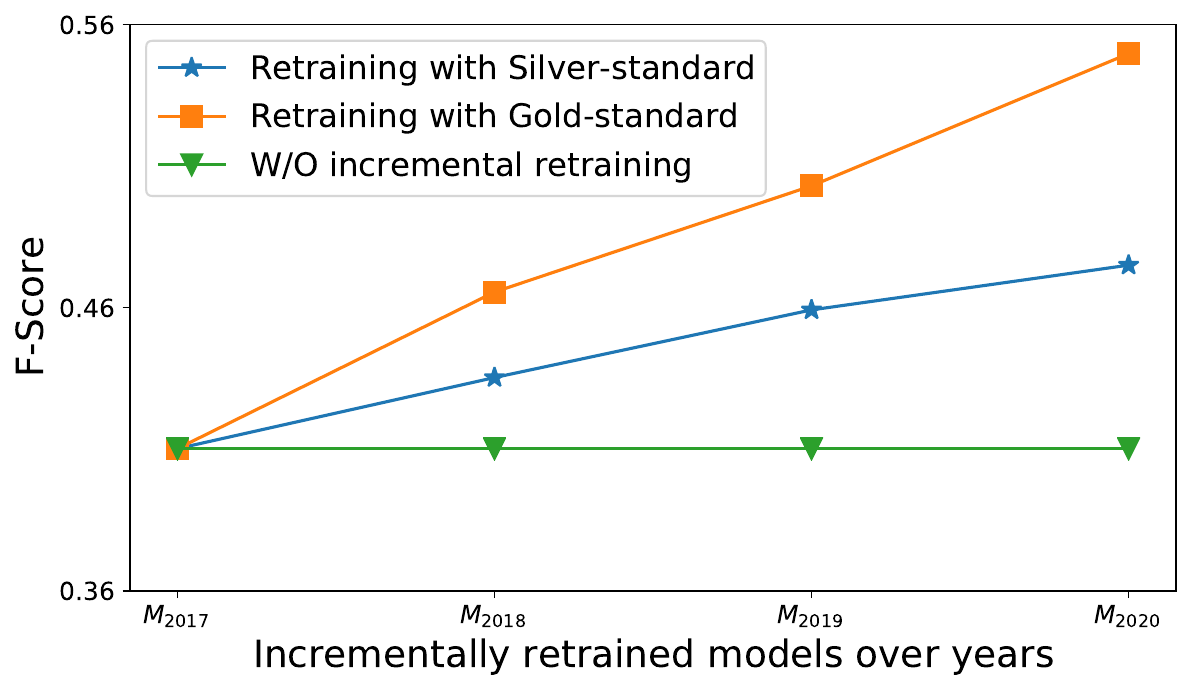} }
    \label{fig:result_rolling_1}
    %}%
    %\qquad
    \subfloat[\centering \small
    %\hlt{
    %Evaluation only on post-2017 entities
    %Performance on post-2017 methods only
    %}
    ]{{\includegraphics[width=.45\textwidth]{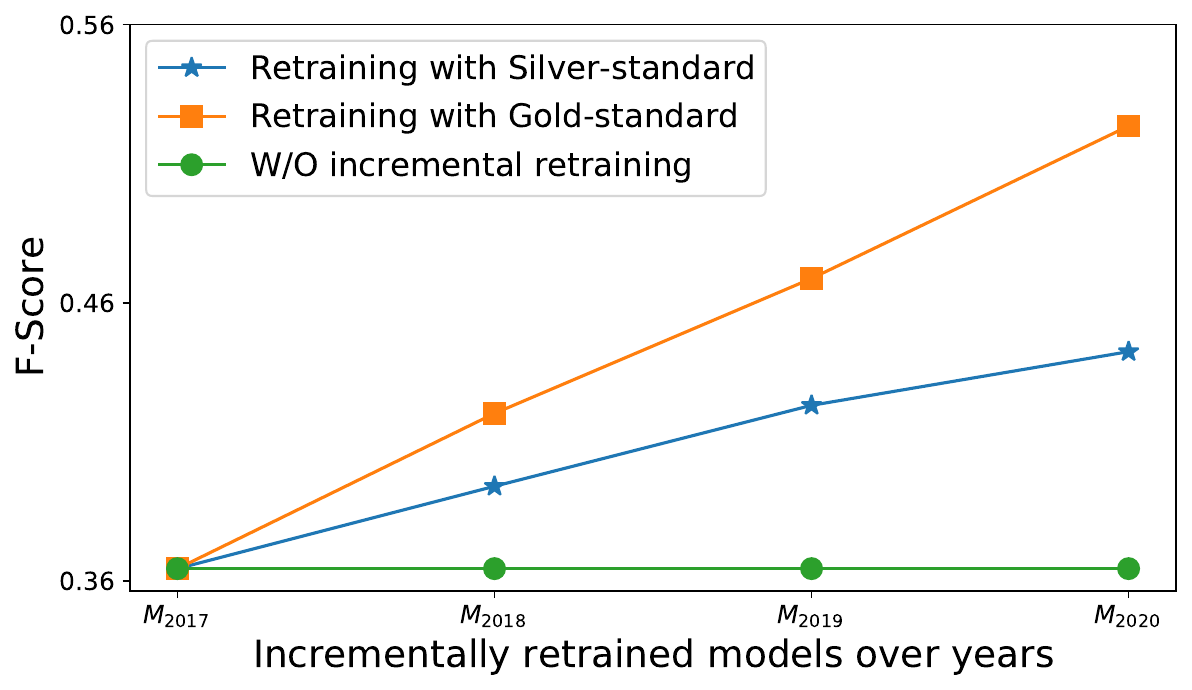} }
    \label{fig:result_rolling_2}
    }
    \caption{\small
    %\hlt{
    Comparisons between the baseline model (M\textsubscript{2017}) and the incrementally retrained models -M\textsubscript{2018}, M\textsubscript{2019} and M\textsubscript{2020}. The models were retrained with silver-standard and gold-standard data (as an apex-line measure). The test set comprises sentences from papers published in 2021 with two different settings - \textit{a}) all entities, and \textit{b}) only on entities appearing after 2017 (zero-shot setup).  %  
    %Note that every model has been tested on data instances from 2021.
    %}
    }
    %\vspace{.6cm}
    \label{fig:result_rolling}
    % \scalebox{0.9}{    \includegraphics[width=1\textwidth,height=.8\textwidth]{Figure/result_line_plot.pdf }}
    % \caption{\small \hlt{Performance of our \FLRFGBinary~ model on chronological time split starting from 2018 in rolling evaluation paradigm.}}
    % \label{fig:result_rolling}
\end{figure}
\begin{figure}[t]%
    \centering
    %\vspace{.6cm}
    \subfloat[\centering \small Performance on pre-2018 and post-2017 methods]{{\includegraphics[width=.45\columnwidth]{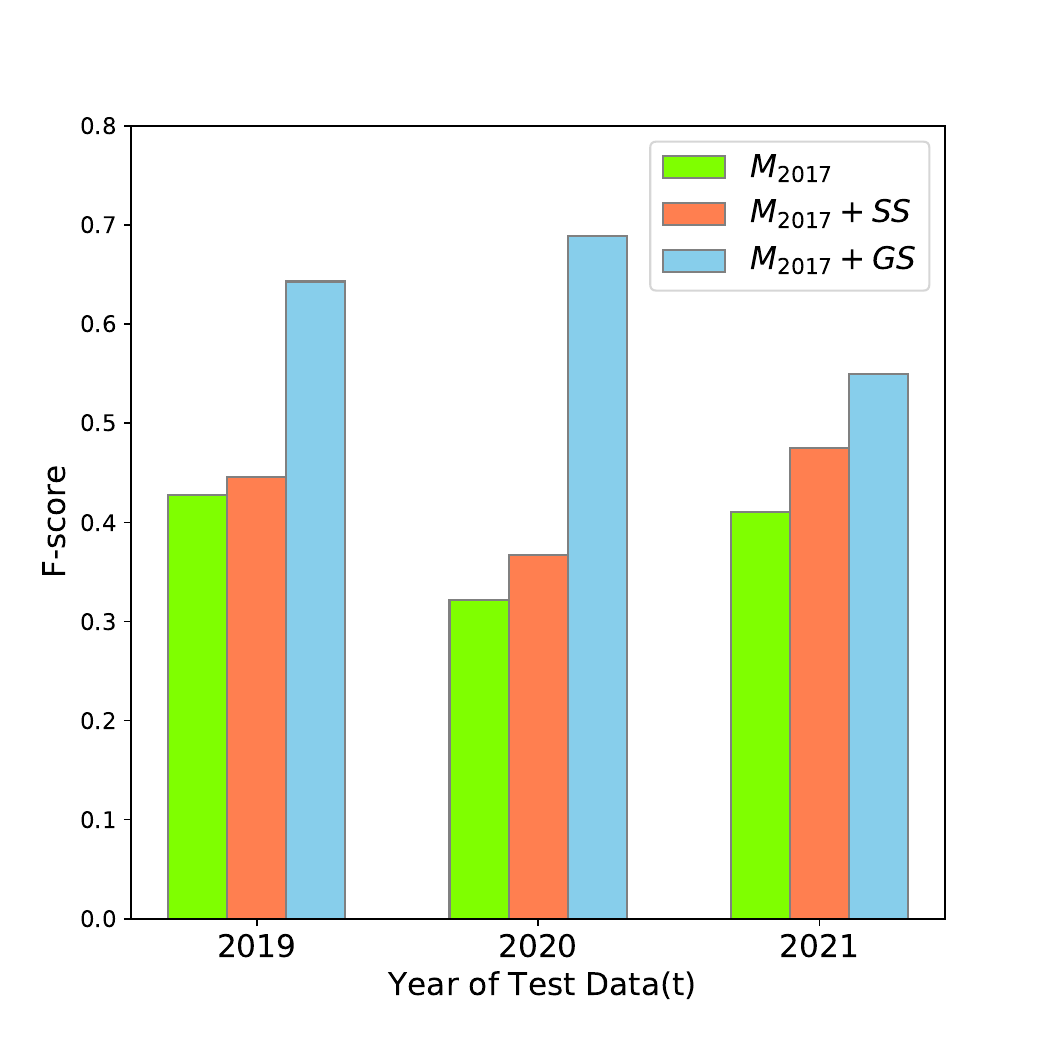} }
    \label{fig:result_rolling_1_n}
    }%
    %\qquad
    \subfloat[\centering \small Performance on post-2017 methods only]{{\includegraphics[width=.45\columnwidth]{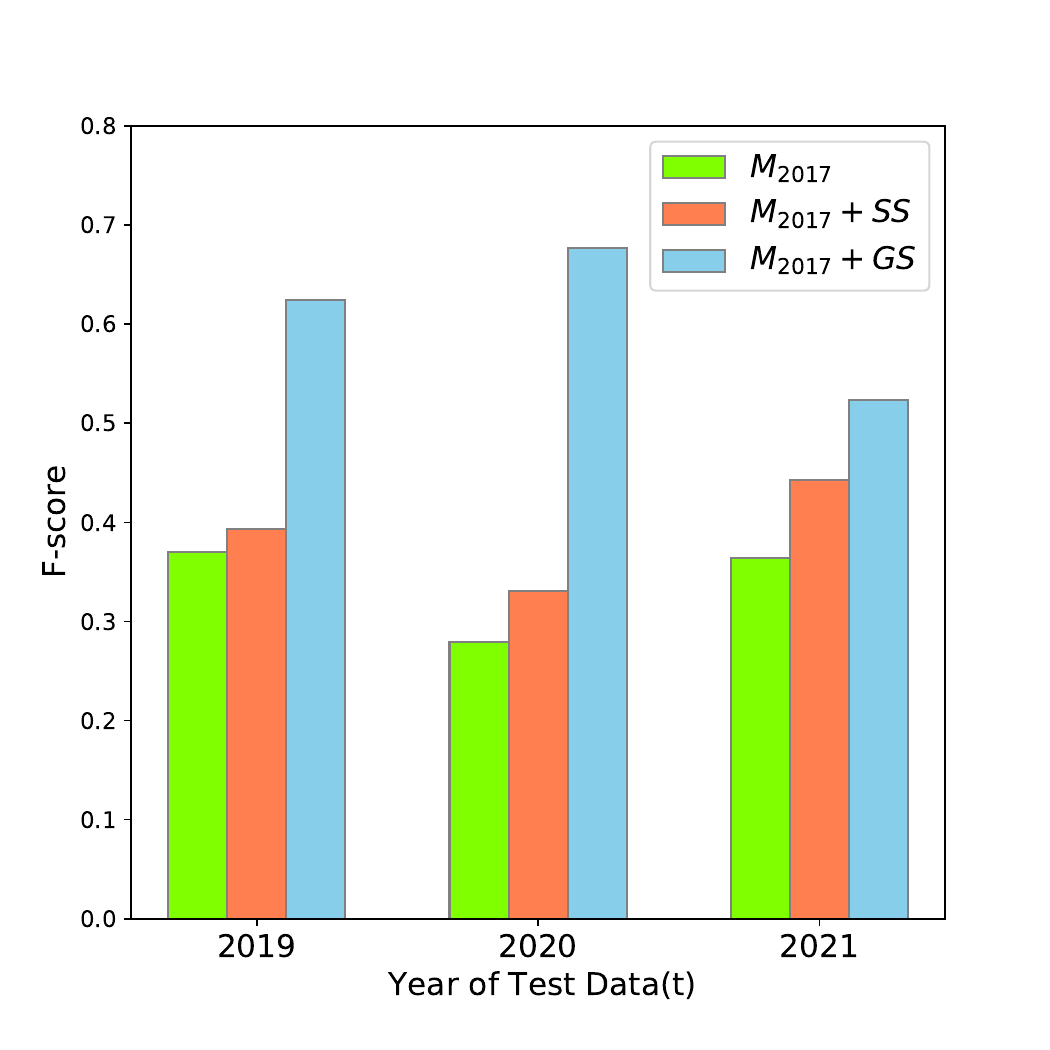} }
    \label{fig:result_rolling_2_n}
    }
    \caption{\small
    %\hlt{
    Similar to Figure \ref{fig:result_rolling}, the difference being here
    we evaluate \FLRFGBinary~at intermediate timestamps, namely years 2019 and 2020 in addition to 2021, the results of which are also included from Figure \ref{fig:result_rolling} for completeness. The notation `SS' corresponds to the case when the predictions from time $t+1$ is fed back into the model instantiated at time $t$ ($M_t$) to yield $M_{t+1}$ (similarly `GS' denotes the case when gold-standard data is used for model updates thus referring to an oracle setup).
    Again similar to Figure \ref{fig:result_rolling}, evaluation is conducted with \textit{a}) all entities, and \textit{b}) only on entities appearing after 2017 (zero-shot setup).    
    %The model at timestamp $t+1$ is updated by using the silver-standard and gold-standard 
    %Performance of \FLRFGBinary~model using M\textsubscript{2017}(trained on pre-2018 data) and incrementally retrained models produced by using gold-standard (GS) and silver-standard (SS) data from consecutive years. Note that the training data (irrespective of whether the model has been incrementally retrained) consists of instances till the end of $(t-1)^{th}$ year and evaluation is on $t^{th}$ year.
    }
    %}
    %\vspace{.6cm}
    \label{fig:result_rolling_second}
    % \scalebox{0.9}{    \includegraphics[width=1\textwidth,height=.8\textwidth]{Figure/result_line_plot.pdf }}
    % \caption{\small \hlt{Performance of our \FLRFGBinary~ model on chronological time split starting from 2018 in rolling evaluation paradigm.}}
    % \label{fig:result_rolling}
\end{figure}

\para{Category-specific analysis}
%\hlt{
Figure~\ref{fig:result_analysis} presents a category specific analysis of the results for all the proposed models including the baselines. A few key observations are as follows.
\uls
\item \FLRFGBinary~consistently performs well across all the 7 different categories.
\item \FLRFGBinary~performs very well on `Audio' domain (cf. Figure~\ref{fig:percat-stats}) in spite of introduction of a significant number of new methodologies in the post-2017 era.
\item Proposed \FLRCGBinary~produces high F-scores across each category. A common intuition behind it is that we develop the \FLRCGBinary~framework by partitioning the input data and that might have enabled it to learn more category specific information from the partitioned data samples.
\ule
These observations suggest that the factored extension of~\FLRCG~is able to generalize well for most scientific disciplines.
%}

 % An interesting observation is that \FGBinary~performs significantly better for the domains SEQ, RL, which are the domains constituting a significant number of new methodologies (see Figure \ref{fig:percat-stats}). Moreover, it also performs the best on the NLP domain.

% \hlt{In response to \textbf{RQ-3}, we adopt incremental retraining strategy and rolling evaluation technique on our proposed \FLRFGBinary~instead of training a new model from scratch over different chronological splits of our data samples. Generally, if we train a model from scratch for each time split the number test instances will reduce drastically for that reason we have adopted retraining technique. Also the application of incremental retraining strategy on an already trained model in a timely manner is crucial in real-world scenario for our downstream new \emph{methodology component} names extraction task. With the aid of this retraining technique, the current model can get more specifically accustomed to the newer context.} 

\subsection{Feedback-based Model Updates}  \label{ss:incremental}

%\textcolor{blue}{

To address RQ-3 we employ the retraining strategy of Section \ref{ss:retrain}, i.e., we use a model's own predictions on the test set as feedback information to incrementally update its parameters. 
In particular, we use the best performing model of Table \ref{tab:outdomain_stat}, namely \FLRFGBinary~for this set of experiments. To gain further insights into the model's behaviour under the zero-shot setup, we partition the set of entities in our test-set (post-2017 as outlined in Section \ref{ss:dataset}) into two groups - one where the timestamp of the first occurrence of the entity precedes 2018, and the other includes only those entities which appear from 2018. Moreover, for an oracle-based comparison we also include results where the models are updated with gold-standard data.

From Figure~\ref{fig:result_rolling} we observe that the model \FLRFGBinary~trained with silver-standard data continually improve over the years (as can be seen from the monotonically increasing values of the middle lines). This line corresponding to the silver-standard setup, as expected, is sandwiched between the two extremes of updates with gold-standard data (the top-line) and no updates (the bottom line). Particularly interesting is the observation for the zero-shot setup (Figure ~\ref{fig:result_rolling_2}), where the middle line is seen to be more closely following the top line, thus indicating the effectiveness of the model updates for new methodology names.

% Now if we consider the time split 2018 where we will have comparison between three models (i.e., M\textsubscript{2017} (baseline one), M\textsubscript{2018} and M'\textsubscript{2018}) and from Figure~\ref{fig:result_rolling_1} and~\ref{fig:result_rolling_2} we can observe that the performance of M'\textsubscript{2018} lies between M\textsubscript{2017} and M\textsubscript{2018} which is very much obvious.

% Additionally, Figure~\ref{fig:result_rolling_1} and~\ref{fig:result_rolling_2} demonstrate the gradual improvement of model performance even trained with the silver-standard data samples, which shows that it is feasible to develop a model for predicting novel methodology names even with the inclusion of silver-standard data. However, we can notice from both Figure~\ref{fig:result_rolling_1} and~\ref{fig:result_rolling_2} that the performance gap between the models trained on GS dataset on different year split and the models trained on SS dataset is increasing rapidly over time because the predictions used to further train the SS model (i.e., the quality of the silver-standard data ) becomes highly unreliable.

%}

%\hlt{
Additionally, to see if this trend is similar across the intermediate years (and not only 2021) we also evaluate the models over the test sets corresponding to the intermediate timestamps (i.e., years 2019 and 2020).
%
%Secondly, we also show the effectiveness of our SS models in different time split. In this scenario, we consider the D\textsubscript{2019} as the required test data instances at the year 2019 and we will have the comparison between M\textsubscript{2017}, M\textsubscript{2018} and M'\textsubscript{2018}. 
Figure~\ref{fig:result_rolling_second} shows that this sandwiching trend is consistent for the intermediate years as well, again indicating the robustness of the feedback method as measured across different test sets. One important point to note is that the chart heights across the different intermediate years are not comparable because the evaluation uses different test sets. 

% presents that the performance of M'\textsubscript{2018} is better than baseline M\textsubscript{2017} but cannot achieve the performance of M\textsubscript{2018}. A obvious reason is that a model trained on SS data instances cannot perform as good as the model trained on GS data samples. Similar observation can be observed from Figure~\ref{fig:result_rolling_1_n} and ~\ref{fig:result_rolling_2_n} for the rest of the years i.e., 2020 and 2021.
%}

% \hlt{These findings suggest to introduce human in the loop atleast after one year to annotate the required data sample manually and retrain the required model to achieve better performance in the downstream task.}

% Each chronological split at $year=y$
% We set up the dataset appropriately in order to carry out the experimental %evaluation in this paradigm.
% %First, we
% involves dividing the entire post-2017 dataset into two different time splits, such as datasets containing only documents from 2018 and 2019 and so forth.
% For this evaluation, we select the best performing model from Table \ref{tab:outdomain_stat}, i.e., \FLRFGBinary.
%In order to calculate prediction performance, we only used the data from 2018.
%We then apply the incremental retraining strategy by training the pretrained model only on 2018 datapoints and computing the prediction performance on 2019 datapoints.
%This will continue until 2021.
%
%\input{tabledefs/example1}
\begin{figure}[t]%
    \centering
    \subfloat[\centering \small
    Historical data (up to 2017)
    %Surrounding context of `transformer' in pre-2018 era
    ]{{\includegraphics[width=.4\columnwidth]{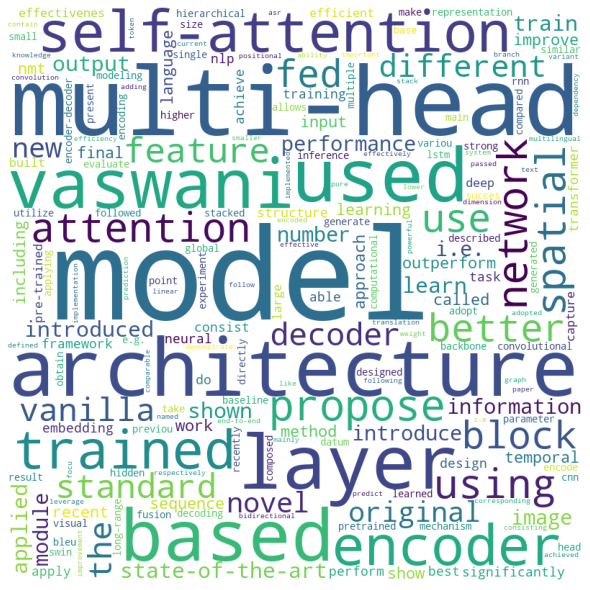} } \label{fig:before_transformer}
    }%
    %\qquad
    \subfloat[\centering \small
    New data (from 2018)
    %Surrounding context of `transformer' in post-2017 era
    ]{{\includegraphics[width=.4\columnwidth]{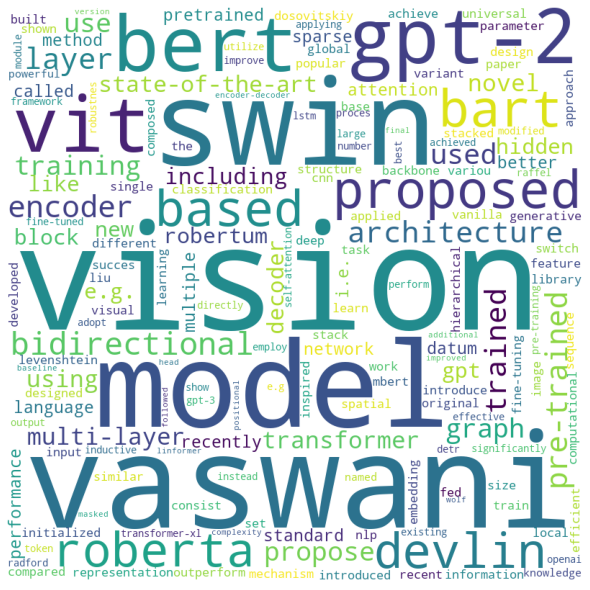} }
    \label{fig:after_transformer}
    }%
    \caption{\small    
    Word cloud illustrating the evolution of the methodology name `transformer' considering its context taken from \textbf{a)} pre-2018 training dataset, and \textbf{b)} model's prediction of `transformer' on post-2017 test dataset.    
    It is to be noted that this word cloud representation is different from the one shown in Figure~\ref{fig:indomain_outdomain} which demonstrated the relative frequencies of the methodology names themselves (as opposed to the context terms of the specific word `transformer').
    }
    \label{fig:evolution}
\end{figure}

%\para{Analyzing our proposed model's ability to capture how the surrounding context changes over time in relation to a methodology name}

\subsection{Additional Analysis}

\para{Changes in contexts of scientific methods}

To illustrate that it is indeed challenging for the models to predict newer entities and to demonstrate the necessity for model updates, we now demonstrate how does the context of a specific word, namely `transformer', changes over time.
While it is seen that the prevalent context terms in the training data (pre-2018 entities) comprises terms such as `architecture', `self-attention', `encoder' etc.,
%`model', `attention', 'multi-head' etc.,
the ones in the test set, on the other hand, 
correspond to `BERT', `VIT', `GPT-2' etc.
%`swin', `roberta', `gpt-2', `bart', `vaswani', `devlin', `model' etc.
This shows that there is a substantial change in the contexts around the mentions of the word `transformer', which in turn shows that firstly, the base model \FLRFGBinary (trained till 2017) performs well in a challenging setup (Table \ref{tab:outdomain_stat}), and secondly, that the feedback with the predicted entities enable the model to capture some of this new context.

\para{Post-hoc manual analysis}
%\hlt{
Additionally, we created a small manually annotated benchmark of data samples from the post-2017 partition of the dataset comprising a total of 285 sentences (again, we will release this as a part of the dataset). We then evaluate the best performing model of Table \ref{tab:outdomain_stat}, namely \FLRFGBinary, with this small manually annotated dataset. 
%
%carry out the prediction performance on this manually annotated \emph{methodology components} using the \FLRFGBinary~model.
We observed that the model yields a precision of 0.623, a recall of 0.2099 and an F-score of 0.314, the numbers being comparable to those reported in Table \ref{tab:outdomain_stat}.
%}.

\subsection{Methodological Implications}
\label{ss:mthd_impl}
\hlt{From the investigations carried out in this work, we observe from Table~\ref{tab:outdomain_stat} that our proposed novel factored-based approach, \FLRFGBinary, achieves state-of-the-art performance compared to the previous model for extracting emerging methodology names from scientific AI literature in our proposed chronological evaluation setup. Furthermore, we have demonstrated that the models trained on historical data can be retrained incrementally using the predictions of a previously trained model for newly published articles. This approach reduces the burden of training a model from scratch and lowers annotation costs. As discussed in Section~\label{ss:incremental}, our incremental training strategy significantly improves the extraction of newer methodology names, which are typically introduced in recent literature.}

\section{Conclusions}
\label{ss:conclsn}
 In this paper we presented a thorough investigation of the feasibility of applying supervised approaches trained on historical data in automatically
extracting novel methodology component names from scientific articles. Focusing specifically on the AI domain, we propose two ways of leveraging the category information of these papers via a factored approach. More specifically, while one of these methods trains an individual model for each category, the other approach constructs more fine-grained labels for training. Our experiments show that both these variants outperform a standard BERT-based fine-tuning approach. While the input data partitioning approach helps increase precision, training with fine-grained labels, on the other hand, leads to increase in the recall and the F-score measures. To the best of our knowledge, this is the first contribution towards extracting emerging methodology names from AI articles in zero-shot and few-shot scenario.

In future, we will expand our work beyond the AI domain to see how the method generalizes to a diverse range of characteristically different domains of study, e.g., economics, physics, etc. We will also explore clustering based approaches to propose a data-driven factored model. Another direction of investigation would be to find out effective ways of filtering the silver-standard data used for incrementally updating the extraction models.

%\st{In future, we seek to improve our proposed approach by incremental retraining, where we plan to utilise weak signals propagated from the predictions on less recent data to further facilitate the prediction on more recent data, e.g., leveraging the information from the predictions of 2018 to help improve the prediction for 2019, and so on. In this work, the automatic evaluation of our model is limited by the silver-standard dataset, which we have prepared. Subsequently, we plan to prepare a gold-standard test dataset.}
% some of the data from 
% In future we will work on extracting evolving tasks, datasets and metrics, and subsequently identifying relations between them, which will pave the way for building an evolving KB in the scientific domain which can track the evolving methodologies, tasks, datasets and metrics. This will ultimately help the research community in tracking progress and keep themselves abreast of the latest research.
%
%\bibliographystyle{splncs04}
\printcredits

\section*{Declaration of Competing Interest}
We do not have any conflict of interests regarding our submission.

\bibliographystyle{cas-model2-names}
\bibliography{main}

 % e.g., identifying the newer sense of the word `transformer' which now usually in most contexts refers to a self-attention based neural architecture

\end{document}